\documentclass[%
 aip,
 amsmath,amssymb,
 reprint,%
]{revtex4-1}

\usepackage{graphicx}
\usepackage{dcolumn}
\usepackage{bm}

\usepackage[utf8]{inputenc}
\usepackage[T1]{fontenc}
\usepackage{mathptmx,comment}
\usepackage{etoolbox,color}

\makeatletter
\def\@email#1#2{%
 \endgroup
 \patchcmd{\titleblock@produce}
  {\frontmatter@RRAPformat}
  {\frontmatter@RRAPformat{\produce@RRAP{*#1\href{mailto:#2}{#2}}}\frontmatter@RRAPformat}
  {}{}
}%
\makeatother
\begin{document}

\preprint{AIP/123-QED}

\title[Mini-review: Phase Transitions in Anisotropic Turbulence]{ Phase Transitions in Anisotropic Turbulence}
\author{Adrian van Kan$^*$}\email{avankan@berkeley.edu}

 \affiliation{Department of Physics, University of California at Berkeley, Berkeley, California 94720, USA}

\date{\today}

\begin{abstract}
Turbulence is a widely observed state of fluid flows, characterized by complex, nonlinear interactions between motions across a broad spectrum of length and time scales. While turbulence is ubiquitous, from teacups to planetary atmospheres, oceans and stars, its manifestations can vary considerably between different physical systems. For instance, three-dimensional (3D) turbulent flows display a forward energy cascade from large to small scales, while in two-dimensional (2D) turbulence, energy cascades from small to large scales. In a given physical system, a transition between such disparate regimes of turbulence can occur when a control parameter reaches a critical value. The behavior of flows close to such transition points, which separate qualitatively distinct \textit{phases} of turbulence, has been found to be unexpectedly rich. 
Here, we survey recent findings on such transitions in highly anisotropic turbulent fluid flows, including turbulence in thin layers and under the influence of rapid rotation. We also review recent work on transitions induced by turbulent fluctuations, such as random reversals and transitions between large-scale vortices and jets, among others. The relevance of these results and their ramifications for future investigations are discussed.

\end{abstract}

\maketitle

\begin{quotation}
The equations of motion of viscous fluids were first formulated 200 years ago by Navier.\cite{navier1827memoire} 
The study of their solutions continues to reveal important new insights, specifically into the intricate properties of turbulence. The last decade has seen significant progress in understanding transitions between qualitatively distinct \textit{phases} of turbulence. Here, we describe part of these recent advances with a focus on highly anisotropic flows relevant, in particular, to geo- and astrophysical applications. This article is an invited mini-review as part of the Dissertation Award in Statistical and Nonlinear Physics of the American Physical Society.\cite{10.1063/5.0181336}
\end{quotation}

\section{\label{sec:level1}Introduction 
}
Werner Heisenberg, whose  foundational contributions to quantum mechanics were recognized with the Nobel Prize in Physics, is said to have uttered on his deathbed that, if he were allowed to ask God two questions, they would be ``Why relativity?'' and ``Why turbulence?'' and he was allegedly sure that God would be able to answer the former question. Turbulence here refers to a state of fluid flows realized when the Reynolds number $\rm Re$ (the ratio of the magnitudes of inertial to viscous forces) attains large values, leading to strongly nonlinear, multi-scale, chaotic dynamics. Although the quote may be apocryphal, Heisenberg did indeed devote significant efforts to the theoretical study 
 of turbulence\cite{heisenberg1985theory} in three spatial dimensions.   
However, his efforts were not crowned with success -- it was Andrey Kolmogorov who, in 1941, proposed the first largely successful attempt at a quantitative theory\cite{kolmogorov1941local} of the \textit{forward energy cascade} transferring kinetic energy from large forcing scales to small, dissipative scales in three-dimensional (3D) turbulence, now understood to be a consequence of nonlinear vortex stretching\cite{frisch1995turbulence} and strain self-amplification.\cite{johnson2020energy,johnson2021role} \\

In the 1950s, there were significant advances in numerical weather prediction,\cite{charney1950numerical} realizing a vision formulated decades earlier by Lewis Fry Richardson.\cite{richardson1922weather} This progress relied on the first available digital computers to solve the barotropic, i.e. two-dimensional (2D), vorticity equation, an idealized model of large-scale motions in the Earth's atmosphere. Around that time, there were first inklings \cite{fjortoft1953changes} that the properties of such turbulence restricted to two spatial dimensions were fundamentally different from its 3D counterpart due to the simultaneous conservation of kinetic energy and enstrophy (mean squared vorticity). The full implications of this observation were first laid out in the seminal contributions of Robert Kraichnan \cite{kraichnan1967inertial,kraichnan1971inertial} (complemented by the works of Leith\cite{leith1968diffusion} and Batchelor\cite{batchelor1969computation}), establishing that turbulence in 2D fluids exhibits a dual cascade, with kinetic energy being transferred towards larger scales (an \textit{inverse energy cascade}), and enstrophy being transferred towards smaller scales (a \textit{forward enstrophy cascade}) from a given stirring scale. Fluid flow in two dimensions is an adequate (albeit often approximate) description for many systems where one spatial dimension is strongly constrained, such as in a thin-layer geometry or in the presence of some other physical mechanism, like rotation or strong magnetic fields, preventing variations in the velocity field along one dimension. This situation is in fact encountered across physical scales from atomic through planetary. At the smallest scales, strongly correlated electron systems in ultra-pure materials such as graphene exhibit 2D fluid-like behavior\cite{levitov2016electron,bandurin2018fluidity, keser2021geometric,narozhny2022hydrodynamic} \textcolor{black}{and while the flows in this context are typically highly viscous and non-turbulent, vortices have been observed in such electron fluids.\cite{aharon2022direct}} At similarly small scales, another striking realization of 2D flow is found in paraxial fluids of light.\cite{ferreira2022towards,baker2023turbulent,congy2024topological} Yet another example of 2D flows is met in Bose-Einstein condensates.\cite{horng2009two,reeves2012classical,seo2017observation,gauthier2019giant,johnstone2019evolution} At larger scales, it has been shown that 2D equations of motion provide a good approximation for flows in thin films of dense bacterial suspensions \cite{wu2000particle,sokolov2007concentration,kurtuldu2011enhancement,wensink2012meso} as well as flows in soap films.\cite{martin1998spectra,rivera1998turbulence,vorobieff1999soap,kellay2002two,shakeel2007decaying} At yet larger scales,  flows in plasmas subject to strong magnetic fields, like those occurring in Tokamak devices, also display a near-2D structure.\cite{xia2003inverse,fujisawa2008review} Even at the large scales of planetary atmospheres, flows are typically constrained to be approximately 2D due to the combined effects of geometric confinement, planetary rotation and density stratification.\cite{byrne2013height,king2015upscale,young2017forward,siegelman2022moist,alexakis2024large} The literature on 2D turbulence is extensive, and this mini-review is not intended to provide an exhaustive description. Instead, we point the reader to existing review articles on the topic.\cite{kraichnan1980two,tabeling2002two,kellay2002two,boffetta2012two}

A noteworthy feature of 2D turbulence is that in a finite system the inverse cascade can cause energy to pile up at the largest scales, thereby producing large-scale Bose condensation, first observed in experiments by 
Sommeria\cite{sommeria1986experimental} and later in simulations by 
Smith and 
Yakhot\cite{smith1993bose} (see also the theoretical work by Falkovich \cite{falkovich1992inverse}), which is associated with the emergence of coherent flow structures such as large-scale vortices or unidirectional jets. The possibility of such condensation had already been suggested earlier by Kraichnan,\cite{kraichnan1967inertial} who proposed the idea that equilibrium statistical mechanics based on energy and enstrophy conservation might be used to describe the condensate, building on related earlier work by Lee \cite{lee1952some} and  Onsager.\cite{onsager1949statistical} The later groundbreaking works by Sommeria \cite{robert1991statistical} and Miller\cite{miller1992statistical} in the 1990s significantly refined these approaches, leading to detailed statistical mechanical descriptions of the large-scale flows based on the 2D Euler equation of ideal (i.e., inviscid and unforced) fluids. This sparked a number of studies on the statistical mechanics of two-dimensional and geophysical flows, reviewed by Bouchet and Venaille.\cite{bouchet2012statistical} Complementing the statistical mechanics approach which relies on the  Euler equations of ideal fluids, a significant body of recent work has established a detailed characterization of forced-dissipative large-scale condensates within 2D turbulence using a combination of direct numerical simulations (DNS) and theoretical approaches.\cite{chertkov2007dynamics,laurie2014universal,frishman2017culmination,frishman2018turbulence,doludenko2021coherent}    \textcolor{black}{We stress that the term \textit{2D} flow is used here to refer specifically to \textit{planar} flow (with two components), which should distinguished from two-dimensional three-component (2D3C) flow, sometimes also referred to as 2.5-dimensional flow, where the velocity field only depends on two coordinates, but retains three components. For instance, even when the flow becomes vertically invariant, e.g., in a very thin layer or under the influence of rapid rotation, the vertical component of velocity obeys a passive scalar equation with an associated forward energy cascade (if the domain has periodic boundary conditions in the vertical). 
A problem for which this distinction is important is the magnetohydrodynamic dynamo instability: 2D3C flows can induce magnetic fields, i.e. dynamo action,\cite{seshasayanan2016turbulent,seshasayanan2017transition} while planar flows cannot.\cite{zeldovich1980magnetic} The transition between 2D flow with two components and 2D3C flow has recently been investigated in a toroidal geometry\cite{Agoua_Favier_Morales_Bos_2024}, but will not be discussed further here. } 

While 2D turbulence has attracted great interest in its own right, in most cases it remains only an approximation for real physical systems. Even when close to 2D, most realistic fluid flows (such as the Earth's atmospheric circulation which is largely confined within the troposphere whose depth of $\sim 10{\rm km}$ is small compared to the horizontal size of storms of $\sim1000{\rm km}$) generically retain some degree of three-dimensionality. Specifically in the atmosphere, an important source of such 3D motions is 
(moist) convection. Their presence motivates the study of the broader class of \textit{quasi-2D} flows, where the fluid motion is constrained to occur \textit{primarily} in two dimensions, but where variations along the third dimension (referred to as the \textit{vertical} in the following) may be present. These vertical variations can strongly impact the dynamics, since they introduce vortex stretching which breaks exact enstrophy conservation, and thus the nature of the resulting turbulent cascades is non-trivial. While quasi-2D turbulence is a problem of long-standing interest, studied in early simulations since the late 1990s \cite{smith1996crossover,celani2010turbulence} (see also the review in Ref.~\cite{danilov2000quasi}), the last few years in particular have seen substantial progress in our understanding of this important class of fluid flows.
In particular, it has been realized that quasi-2D turbulence features a \textit{bidirectional} (or \textit{split}) \textit{energy cascade}, where energy injected at some intermediate forcing scale simultaneously cascades both to larger and to smaller scales at rates depending on physical control parameters, e.g. the depth of a thin fluid layer, the planetary rotation rate or the strength of density stratification. A bidirectional cascade in an anisotropic flow is facilitated by the multi-scale nature of turbulence: large-scales can be strongly constrained to exhibit approximately 2D dynamics, while small scales are unrestrained, evolving as fully 3D. \textcolor{black}{The upscale transfer of kinetic energy is observed in tropical cyclone atmospheric boundary layer dynamics\cite{shao2023impact}, where it leads to reduced drag, which can lead to forecast errors if not parameterized correctly.}

Given the starkly contrasting phenomenology of 2D, 3D and quasi-2D turbulence, respectively, it is natural to investigate the possibility of transitions between these disparate regimes as physical control parameters 
are varied. Such transitions are quantified in terms of an order parameter $O$ such as the inverse energy flux or the large-scale kinetic energy as a function of a control parameter $\mu$, e.g., layer height. In general, they can be smooth or occur at a \textit{critical point} (or critical hypersurface in a higher-dimensional parameter space).
If the transition is critical (i.e., occurs at a critical point), then this allows one to distinguish well-defined \textit{phases} of turbulence. A series of recent works has been concerned with the identification and characterization of such transitions in different turbulent flows, which has led to remarkable findings, including several examples of critical points separating forward and bidirectional energy cascades,\cite{seshasayanan2014edge,benavides2017critical,Ecke_2017,van2020critical} and the observation of noise-induced transitions between large-scale condensates and small-scale turbulence,\cite{yokoyama2017hysteretic,van2019rare,de2022bistability} among others. The primary topic of this review is to survey these advances in our understanding of such  transitions in highly anisotropic flows, which rely to a large extent on extensive numerical simulations. We will give a phenomenological description of the different phases of turbulent flows that are observed and describe their rich behavior close to the critical points, as well as the limitations of existing studies and important open questions. Throughout the text, the mathematical formalism is made as light as possible in favor of readability. References are given for more formal descriptions.

We emphasize that this review is focused on transitions between distinct \textit{turbulent} flow states, with energy transfers across a wide range of scales (implying that the system is out of statistical equilibrium) on both sides of the transition. This problem is qualitatively distinct from the transition between \textit{laminar flow} (which does not feature energy fluxes across scales) and \textit{turbulence}. 
This mini-review reflects a subjective choice of topics that the author deems of the greatest interest. Alternative and complementary perspectives are presented in other recently published reviews.\cite{alexakis2018cascades,alexakis2023quasi,boffetta2023dimensional,marston2023recent} 
The remainder of this paper is structured as follows. In Section \ref{sec:bg}, we give background information on the basic properties of 2D and 3D turbulence. In Section \ref{sec:dimensional_transitions}, we summarize recent findings on `dimensional' transitions between 3D, quasi-2D and 2D turbulence in different physical systems, with an emphasis on geo- and astrophysically relevant effects including thin-layer geometry, rotation and density stratification. In Section~\ref{sec:fluctuation_induced_transitions}, we describe examples of anisotropic flows displaying multistability and transitions between distinct turbulent flow states induced by turbulent fluctuations. Next, in Section~\ref{sec:other_transitions_2D_turbulence}, we discuss phase transitions recently identified within variants of 2D turbulence and finally, in Section~\ref{sec:conclusions}, we conclude by putting these developments in a broader context.
\section{Background: turbulence in two and three dimensions\label{sec:bg}}
Physics in two dimensions holds many surprises and turbulence is no exception. Here we provide a brief description of the properties of 3D and 2D turbulence to lay the necessary groundwork for the remainder of the text. For a more complete presentation, we refer the reader to the excellent textbooks on 3D turbulence,\cite{frisch1995turbulence,pope2001turbulent,davidson2015turbulence} as well as existing reviews on 2D, 3D and quasi-2D  turbulence.\cite{tabeling2002two,boffetta2012two,pandit2017overview,alexakis2018cascades,zhou2021turbulence}
\subsection{Three-dimensional turbulence\label{sec:bg_3D}}
Turbulence in a 3D fluid of constant density can be described (in its most basic form) by the evolution of the velocity field $\mathbf{u}$ following the incompressible Navier-Stokes equation
\begin{equation}
    \partial_t \mathbf{u} + \mathbf{u}\cdot \nabla \mathbf{u} = - \nabla P + \nu \nabla^2 \mathbf{u} - \alpha \overline{\mathbf{u}} + \mathbf{f}, \qquad \nabla\cdot \mathbf{u} = 0,\label{eq:nse}
\end{equation}
 where $P$ is the pressure divided by the constant density, $\nu$ is the kinematic viscosity and $\mathbf{f}$ is an externally imposed body force injecting power $\epsilon$ on a well-defined forcing scale $\ell_f=2\pi/k_f$ (analogous to the size of a stirring device), with the corresponding forcing wave number $k_f$. The term $-\alpha \overline{\mathbf{u}}$, where $\overline{(\cdot)}$ denotes the vertical average, is an additional ``bottom drag'' modeling the effects of boundaries. 
 We consider a Cartesian domain of dimensions $L\times L \times H$, subject to \textcolor{black}{triply} periodic boundary conditions for simplicity, where $H$ (assumed to be less than $L$) is varied to study the suppression of 3D variations in the flow. \textcolor{black}{All results discussed in the following assume periodicity in the vertical direction, unless otherwise stated. Alternative, e.g., no-slip boundary conditions lead to the formation of boundary layers which may impact the flow, but this is not considered here.}

The idealized setup defined above is attractive due to its simplicity. For a given choice of the forcing function, it is fully characterized by four non-dimensional parameters: 
\begin{itemize}
    \item[(i)] Reynolds number ${\rm Re} = \epsilon^{1/3}/(k_f^{4/3}\nu)$ based on the forcing parameters,
    \item[(ii)] Frictional Reynolds number ${\rm Re_\alpha} = \epsilon^{1/3}k_f^{2/3}/\alpha$,
    \item[(iii)] Aspect ratio $H/L$,
    \item[(iv)] Nondimensional layer height $k_f H$.
\end{itemize}

For smooth, inviscid ($\alpha=\nu=0$) and unforced ($\epsilon=0$) flows, Eq.~(\ref{eq:nse}) conserves 
two quadratic invariants:  the energy $\mathcal{E} = \langle \frac{1}{2}|\mathbf{u}|^2\rangle$, where $\langle \cdot \rangle$ 
represents the volume average, and the helicity $\mathcal{H} =\langle \mathbf{u}\cdot {\boldsymbol{\omega}}\rangle$, where $\boldsymbol{\omega}=\nabla\times\mathbf{u}$ is the vorticity. The role of $H$ has been described elsewhere\cite{alexakis2018cascades,pouquet2022helical} and will not be discussed here. When forcing and dissipation are present,  then in steady state
\begin{equation}
    \epsilon = D^{\mathcal{E}}_\nu + D^\mathcal{E}_\alpha,\label{eq:energy_balance}
\end{equation}
where the time and volume-averaged energy injection rate $\epsilon = \langle \mathbf{f}\cdot \mathbf{u}\rangle$ balances the dissipation rates by viscosity $D^\mathcal{E}_\nu = \nu\langle |\boldsymbol{\omega}|^2\rangle$ and bottom drag $D^\mathcal{E}_\alpha = \alpha \langle|\overline{\mathbf{u}}|^2\rangle$.

A common way to quantify length scales in a continuous, periodic field is in terms of the inverse of the wave number $k=|\mathbf{k}|$ when computing Fourier transforms
\begin{equation}
\hat{\mathbf{u}}_\mathbf{k} = \langle e^{-i\mathbf{k}\cdot \mathbf{x}} \mathbf{u}\rangle, \hspace{0.5cm} \mathbf{u} = \sum_\mathbf{k} \hat{\mathbf{u}}_\mathbf{k} e^{i\mathbf{k}\cdot \mathbf{x}}.
\end{equation}
The (spherically averaged) energy spectrum is then defined as 
\begin{equation}
    E(k) = \frac{1}{2 \Delta k}\sum_{\stackrel{\mathbf{p}}{  k\leq |\mathbf{p}|  \leq k+\Delta k}}|\hat{\mathbf{u}}_\mathbf{p}|^2,
\end{equation}
where $\Delta k = 2\pi/L$ is the minimum wave number present in the system. The spectrum is related to the total energy by $\mathcal{E}=\sum_k E(k)\Delta k$, with $E(k)\Delta k$ being the energy contained in the wave number interval $[k,k+\Delta k]$. \textcolor{black}{Based on the fraction of energy $p_k = E(k)\Delta k/\mathcal{E}$ contained in each wave number shell, Verma al.\cite{verma2022hydrodynamic,verma2024contrasting} recently introduced hydrodynamic entropy to quantify the multiscale disorder of turbulence (not discussed further here).}

\begin{figure}
    \centering
    \includegraphics[width=0.6\linewidth]{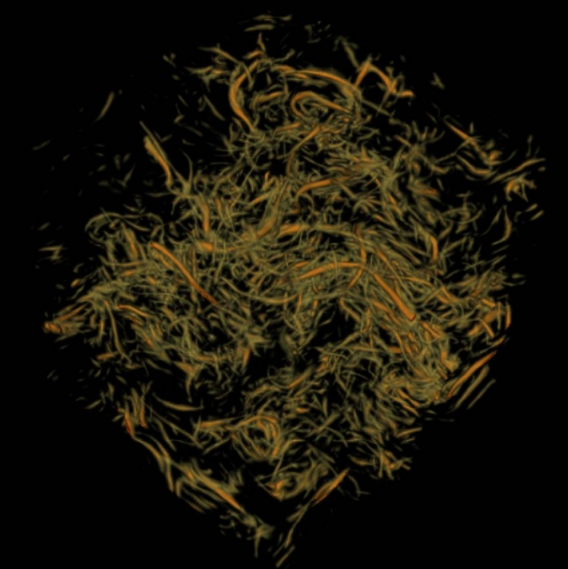}
    \caption{Iso-contours of vorticity from a high-resolution simulation of 3D homogeneous and isotropic turbulence. The elongated small-scale vorticity filaments visible in the snapshot result from vortex stretching.  \textcolor{black}{Source: Ryan McKeown et al. New J. Phys. 2023, 25 103029; licensed under a Creative Commons Attribution (CC BY)
license.}}
    \label{fig:3D_turbulence_vorticity_isocontours}
\end{figure}

The key insight from the first half of the 20th century was that the flux of energy across scales controls the statistical properties of high $\rm Re$ flows. In 1941, Kolmogorov argued\cite{kolmogorov1941local} that in a forward energy cascade all energy is dissipated at small scales and none at large scales, i.e.
 $D_\nu^\mathcal{E} = \epsilon$ and $D^\mathcal{E}_\alpha=0$ in our notation, provided the limits $\rm Re\to \infty$ and $\rm Re_\alpha\to \infty$ are taken. Assuming locality in scale, Kolmogorov predicted that
 \begin{equation}
     E(k) = C_K \epsilon^{2/3} k^{-5/3} \, \quad \text{ for  }\,\,\,\,\,k_f<k<k_\nu,
 \end{equation}
 up to a cut-off wave number (now named after Kolmogorov) where viscous dissipation terminates the forward energy cascade, given by $k_\nu = \epsilon^{1/4} /\nu^{3/4} = k_f {\rm Re}^{3/4}$, and $C_K$ is the Kolmogorov constant. At wave numbers smaller than $k_f$ (corresponding to scales larger than $\ell_f$), 
there is no flux of energy, and thus these scales are expected to reach a thermal equilibrium 
state characterized by an equipartition of energy among modes.\cite{dallas2015statistical,alexakis2019thermal,gorce2022statistical} Later work revealed \textcolor{black}{\cite{frisch1995turbulence}} that the forward energy cascade is not local in scale, and the probability distribution of velocity increments between different points in space is non-Gaussian, developing heavy tails at small scales. This phenomenon, which is known as \textit{intermittency}, is associated with a small deviation from the $-5/3$ spectral exponent and has received much attention, but will not be discussed further here. A key physical mechanism facilitating the forward energy cascade of 3D turbulence is \textit{vortex stretching},\cite{tennekes1972first} i.e. vorticity amplification at smaller scales by along-vortex velocity gradients, which is associated with the term $\boldsymbol{\omega}\cdot \nabla\mathbf{u}$ in the evolution equation for vorticity $\boldsymbol{\omega}$, in addition to strain amplification.\cite{johnson2020energy,johnson2021role} Vortex stretching leads to a small-scale filamentary structure which can be seen in the vorticity iso-contours from a high-resolution 3D homogeneous and isotropic turbulence simulation, shown in Fig.~\ref{fig:3D_turbulence_vorticity_isocontours}.

\subsection{Two-dimensional turbulence\label{subsec:2D_turbulence}}
Fluid flows in a 2D flatland (here taken to be in the $x-y$ plane) also obey the Navier-Stokes equation Eq.~(\ref{eq:nse}), but their dynamics differ markedly from their 3D counterparts. This can be deduced from the evolution of vertical vorticity $\omega_z \equiv\boldsymbol{\hat{z}}\cdot \boldsymbol{\omega} $, where $\boldsymbol{\hat{z}}$ is the unit vector in the $z$-direction, which for 2D flows reads
\begin{equation}
    \partial_t \omega_z + \mathbf{u}\cdot\nabla \omega_z = - \alpha \omega_z+ \nu \nabla^2 \omega_z + f_{\omega_z}, \label{eq:nse2D}
\end{equation}
where $f_{\omega_z}=\boldsymbol{\hat{z}}\cdot (\nabla\times \mathbf{f})$. Note that there is no vortex stretching term on the right-hand side of Eq.~(\ref{eq:nse2D}) since the vorticity and velocity vectors are perpendicular for \textcolor{black}{planar} 2D flow. This means that a key mechanism responsible for the forward energy cascade is absent, with profound implications for energy transfers. Equation~(\ref{eq:nse2D}) has two ideal quadratic invariants: energy $E=\langle \frac{1}{2} |
\mathbf{u}|^2\rangle$ and enstrophy $\Omega = \langle \omega_z^2\rangle$. Moreover, the \textit{Casimir} invariants $\langle g(\omega_z)\rangle$ are conserved for any differentiable function $g$, which presents a challenge in particular for statistical mechanics approaches.\cite{bouchet2012statistical} In a statistically steady state, dissipation balances the injection of energy and enstrophy i.e. 
\begin{equation}
    \epsilon = D^\mathcal{E}_\alpha + D^\mathcal{E}_\nu, \qquad \eta \equiv k_f^2 \epsilon = D_\alpha^\Omega + D_\nu^\Omega,
\end{equation}
where $\epsilon$ is again power injected at the forcing scale, associated with an enstrophy injection rate $\eta$, the energy dissipation rates are given by $D^\mathcal{E}_\alpha= \alpha \langle \mathbf{u}^2\rangle$, $D^\mathcal{E}_\nu = \nu \langle \omega_z^2 \rangle$ and the enstrophy dissipation rates by $D^\Omega_\alpha = \alpha \langle \omega_z^2 \rangle $ and $D_\nu^\Omega = \nu \langle |\nabla\omega_z|^2 \rangle$.

Following his involvement\cite{charney1950numerical} in early numerical weather forecasting based on 2D models, Fj\o{}rtoft\cite{fjortoft1953changes} proposed an argument indicating that kinetic energy cannot be transferred from large to small scales in 2D and must instead be transferred inversely, i.e. from small to large scales. His argument is based on the observation that the enstrophy spectrum $\Omega(k)$ is directly related to the energy spectrum by $\Omega(k)=2k^2E(k)$. Fj\o{}rtoft's argument considers an ideal fluid and assumes that energy is initially concentrated in some wave number interval of width $\delta k = \left(\mathcal{E}^{-1} \sum_k (k-\overline{k}_\mathcal{E})^2  E(k)\Delta k\right)^{1/2}$, where $\overline{k}_\mathcal{E}=\mathcal{E}^{-1} \sum_k kE(k)\Delta k$ is the energy centroid. It can easily be shown that $(\delta k)^2 = \Omega/(2\mathcal{E})- \overline{k}_\mathcal{E}^2$. Since $\Omega/(2\mathcal{E})$ is conserved and nonlinear interactions tend to increase $\delta k$, the energy centroid $\overline{k}_\mathcal{E}$ must decrease, i.e. energy is transferred to larger scales. Similarly defining the enstrophy centroid as $\overline{k}_\Omega=\Omega^{-1}\sum_k k\Omega(k)\Delta k $, it follows (using the Cauchy-Schwarz inequality) that $\overline{k}_\mathcal{E}\overline{k}_\Omega \geq \Omega/(2\mathcal{E})$. This implies that as $\overline{k}_\mathcal{E}$ decreases,  $\overline{k}_\Omega$ necessarily increases, indicating a forward transfer of enstrophy to smaller scales. The resulting balance relations for a dual inverse energy forward enstrophy cascade, in an infinite domain and in the limit of large $\rm Re$, $\rm Re_\alpha$, read
\begin{equation}
     D^\mathcal{E}_\alpha = \epsilon, \,\,\,\, D_\nu^\mathcal{E} =  0 \qquad \text{ and} \qquad D_\alpha^\Omega =0, \,\,\,\,   D_\nu^\Omega = \eta = k_f^2 \epsilon.
\end{equation}
This dual cascade picture deduced from Fj\o{}rtoft's argument, and later fully developed by Kraichnan,\cite{kraichnan1967inertial,kraichnan1971inertial} has been corroborated in high-resolution numerical simulations.\cite{boffetta2007energy,boffetta2010evidence} By similar arguments as in the 3D case, one finds that in 2D the associated energy spectrum at $k_\alpha\ll k\ll k_f$ (in the inverse energy cascade range) is of the form $E(k) \propto k^{-5/3}$, while at $k_f \ll k \ll k_\nu$ (in the forward enstrophy cascade range) it is steeper $E(k)\propto k^{-3}$. The wave number of frictional arrest of the inverse energy cascade is given by $k_\alpha=\alpha^{3/2}/\epsilon^{1/2}
= k_f {\rm Re_\alpha}^{-3/2}$, while the wave number of viscous enstrophy dissipation reads $k_\nu = \eta^{1/6}
/\nu^{1/2} = k_f {\rm Re}^{1/2}$. Note that the $\rm Re$-dependence of $k_\nu$ is different in 2D and 3D.

\begin{figure}
    \centering
    \includegraphics[width=0.9\linewidth]{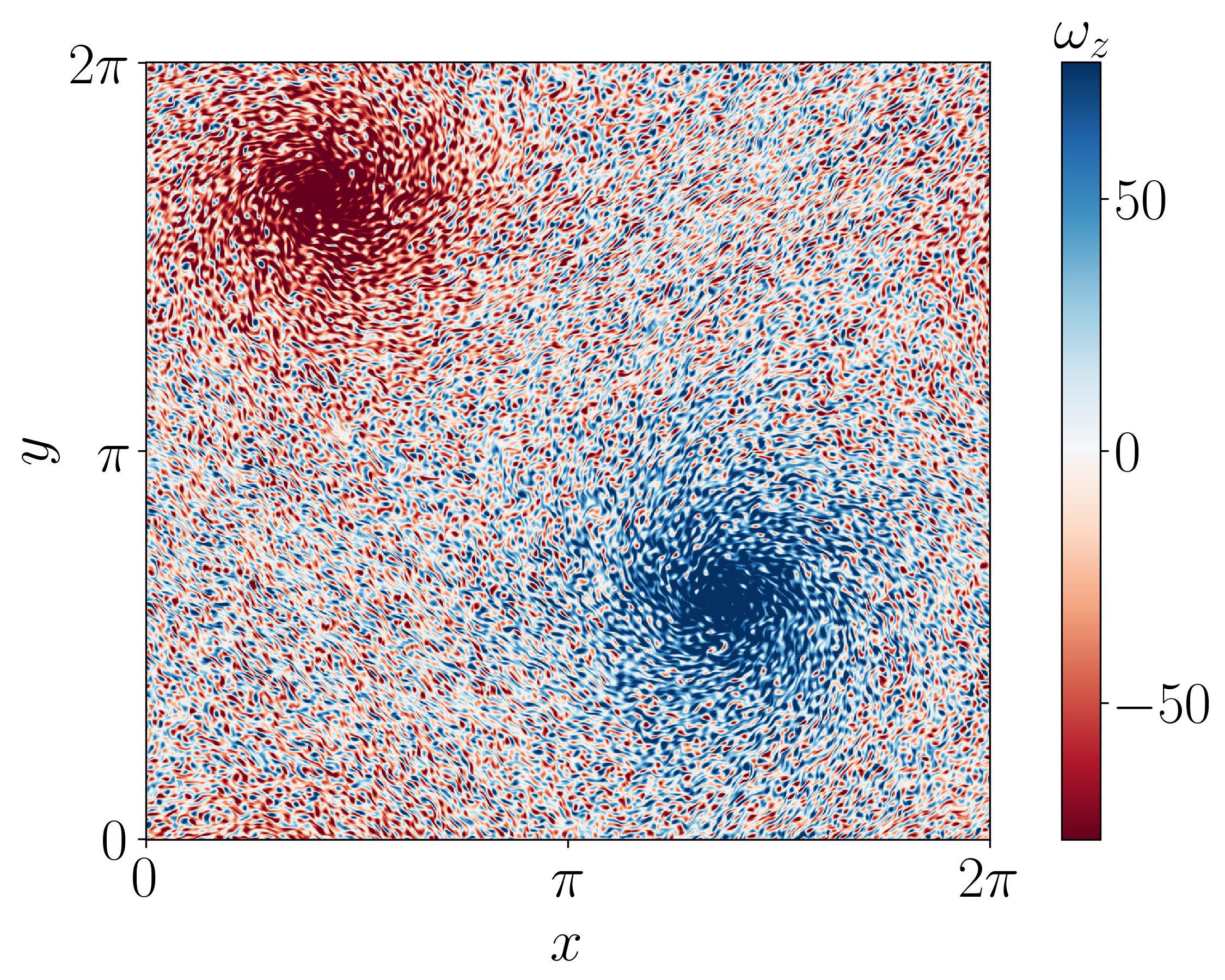}
    \caption{Vorticity field of a large-scale vortex dipole condensate in 2D turbulence obtained at late times from small-scale stochastic forcing in a square periodic domain of size $L \ll L_\alpha$.}
\label{fig:2D_turbulence_vortex_condensate}
\end{figure}

In a finite domain of size $L$, two important cases must be distinguished in terms of $L_\alpha= 2\pi/k_\alpha$: if $L\gg L_\alpha$, then the energy cascades develop as described above, up to the scale of frictional arrest. On the other hand, if $L\ll L_\alpha$, then the inverse energy cascade is arrested at the largest available length scale in the system, i.e. $L$, and energy piles up there to form a condensate. In a square periodic domain, the condensate takes the form of a counter-rotating vortex dipole at the scale of the domain, as illustrated in Fig.~\ref{fig:2D_turbulence_vortex_condensate}.  In the condensate state at large $\rm Re$ and $\rm Re_\alpha$, energy is primarily dissipated at the largest scale, which leads to the following estimate for the energy in the statistically stationary state
\begin{equation}
    \mathcal{E} \propto \frac{\epsilon}{\alpha+4\pi^2 \nu L^{-2}}.
\end{equation}
This relation indicates that for weak drag and small viscosity, the energy can reach very large values.

\section{Dimensional transitions\label{sec:dimensional_transitions}}
In view of the strikingly disparate properties of turbulence in two and three spatial dimensions, it is natural to examine what has been termed the \textit{dimensional transition}\cite{deusebio2014dimensional,boffetta2023dimensional} between these two cases as the degree of anisotropy is increased starting from an isotropic 3D turbulent flow, e.g., by flattening the domain geometry. It is interesting to note in this context that in three dimensions, since the only quadratic invariant in addition to energy is helicity (cf. Sec.~\ref{sec:bg_3D}), which is sign indefinite and not related to the energy, there is no Fj\o{}rtoft-type argument analogous to the 2D case. This indicates that, although in \textit{homogeneous and isotropic} 3D turbulence energy does cascade to small scales, the same is not necessarily true in strongly anisotropic 3D flows. Indeed, the dimensional transition typically occurs in two steps: first, as the flow becomes quasi-2D, the forward energy cascade splits up, with some energy cascading to large scales, while 3D variations persist in the flow, transferring the remainder of the energy to small scales. Provided $L\ll L_\alpha$, even a small inverse energy flux  can  lead to condensation at late times, thus drastically altering the flow. Second, for very strong anisotropy, 3D variations decay and exact 2D flow is obtained (forcing and boundary conditions permitting). 

Below, we briefly summarize important known results on both of these steps of the dimensional transition from 3D to 2D turbulence in different systems. \textcolor{black}{The following section is inspired by the work of Alexakis\cite{alexakis2023quasi} with the permission of the author.} 
\subsection{Thin-layer turbulence}
Large-scale atmospheric and oceanic flows are highly constrained by geometry: they feature horizontal scales of the order of $1000\,$km, but are confined in the vertical to a height of $10\,$km or less. In addition to this geophysical motivation, turbulence confined within thin fluid layers is arguably the simplest system to display split energy cascades. For these reasons, it has been extensively studied. Homogeneous turbulence ($L \gg L_\alpha$) is qualitatively different from the case with a large-scale condensate ($L\ll L_\alpha$). Therefore, we discuss the two cases separately and adopt different order parameters to quantify the dimensional transition.

In homogeneous turbulence, i.e., in the absence of a condensate, it is adequate to directly measure the forward and inverse energy fluxes based on  $\tilde{D}_\alpha^\mathcal{E}= D_\alpha^\mathcal{E}/\epsilon$ and $\tilde{D}_\nu^\mathcal{E} = D_\nu^\mathcal{E}/\epsilon$, corresponding to the \textit{fractions} of energy cascading to large and small scales, respectively. We remind the reader that in the limit of infinite $\rm Re,\, Re_\alpha$, 3D turbulence features $\tilde{D}_\alpha^\mathcal{E} = 0$, $\tilde{D}_\nu^\mathcal{E}=1$, while in 2D turbulence $\tilde{D}_\alpha^{\mathcal{E}}=1$, $\tilde{D}_\nu^{\mathcal{E}}=0$. In the presence of a condensate, it is more appropriate to directly measure the condensate amplitude based on the energy of the largest-scale modes 
\begin{equation}
    \mathcal{E}_{LS} = \sum\limits_{k<k_c} E(k)\Delta k, 
\end{equation}
where $k_c=2n \pi/L$ with $n$ of order one. In addition, to quantify the degree of three-dimensionality of the flow, it is advantageous to measure the kinetic energy contained in 3D modes
\begin{equation}
    \mathcal{E}_{3D} =  \frac{1}{2} \langle |\mathbf{u} - \overline{\mathbf{u}}|^2\rangle,
\end{equation}
where $\overline{(\cdot)}$ represents the vertical average. \textcolor{black}{Measures of chaos based on Lyapunov exponents provide an alternative approach to studying the transition between 2D and quasi-2D turbulence\cite{clark2021chaotic,ho2024chaotic}, but this will not be discussed here. }
\subsubsection{Homogeneous  quasi-2D turbulence}
The first study showing that energy can cascade simultaneously to large and small scales in turbulence confined within thin layers is by Smith et al.,\cite{smith1996crossover} and a number of more detailed subsequent works later quantified energy fluxes as a function of the layer height.\cite{celani2010turbulence,musacchio2017split,benavides2017critical,van2019condensates,poujol2020role} Below, based on these works, we describe how  $D_\nu^\mathcal{E}, D_\alpha^\mathcal{E}$ $\mathcal{E}_{LS}$ and $\mathcal{E}_{3D}$ change
as the layer height $H$ is varied. 

We begin by considering a layer of height $H$ much larger than the forcing scale $\ell_f$, and then gradually reduce $H$. For $H\gg\ell_f$, the flow displays 3D turbulence with a fully forward cascade observed at scales smaller than $\ell_f$, leading to a $-5/3$ power law range in the energy spectrum. This behavior is altered once the layer height is reduced to a threshold height $H_{3D}$ proportional to $\ell_f$. For $H<H_{3D}$, a new phase of turbulence with a bidirectional energy cascade appears. 
\begin{figure}
    \centering
    \includegraphics[width=\linewidth]{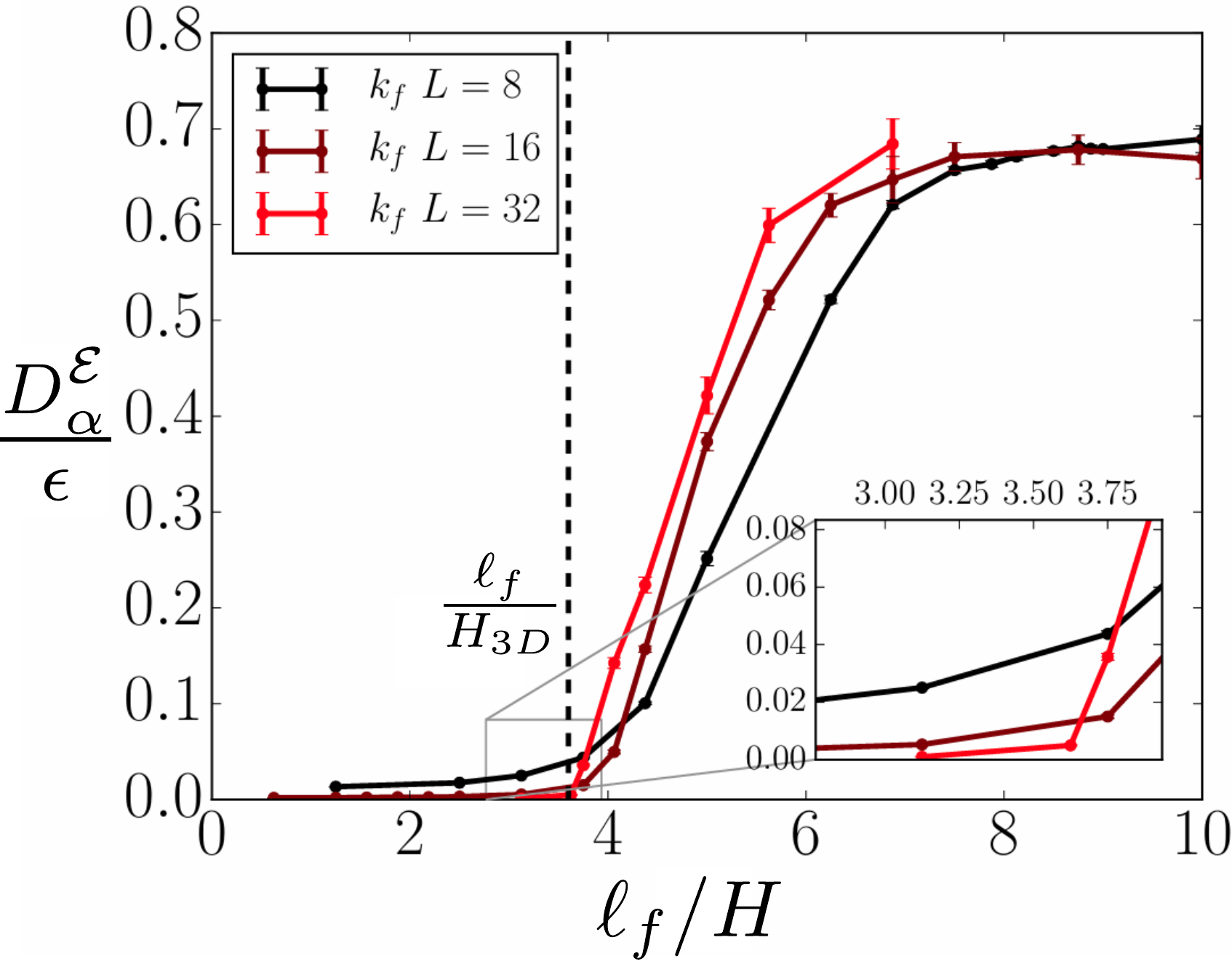}
    \caption{\textcolor{black}{Fraction of injected energy cascading to large scales for turbulence within thin a layer of height $H$ and horizontal length $L$ forced at scale $\ell_f = 2\pi/k_f$. For sufficiently large scale separation between forcing scale and domain size, the critical scaling of the energy flux is approximately linear. Reproduced with permission from  J. Fluid Mech. 822, 364--385 (2017). Copyright 2017 Cambridge University Press.}}
    \label{fig:epsilon_inverse_thin_layer}
\end{figure}
In the absence of a condensate, direct numerical simulations indicate that at $H<H_{3D}$ the fraction of energy cascading inversely increases as
\begin{align}
    D_\alpha^\mathcal{E} \propto (H_{3D}-H)^{\beta_1},
\end{align}
where an exponent $\beta_1$ close to unity is observed, \textcolor{black}{provided a large scale separation between the forcing scale $\ell_f$ and the horizontal domain size $L$} \textcolor{black}{(see Fig.~\ref{fig:epsilon_inverse_thin_layer})} but more work is required to measure it more precisely and complement the existing numerical evidence \textcolor{black}{for free slip\cite{benavides2017critical} and periodic\cite{van2019condensates} boundary conditions in the vertical direction.} No theory exists to describe the physics of this non-equilibrium transition, and therefore more detailed numerical and theoretical studies will be needed to clarify its properties and determine whether it may potentially be related to a known universality class. \textcolor{black}{Recent results derived using renormalization group methods from field theory indicate the presence of a (fractional) cross-over dimension $d\approx2.15$ where the energy flux changes sign, but these results are derived based on assumptions such as a spatially infinite domain, by contrast with the finite-size setup discussed here, and more work remains to be done to examine possible connections.\cite{verma2024critical}}  The very existence of a critical point at $H=H_{3D}$ is nontrivial, and remains to be further investigated, given that even bifurcations in extremely well-controlled low-dimensional dynamical systems show some rounding due to experimental imperfections.  While the available evidence\cite{poujol2020role} indicates that the existence of $H_{3D}$ is robust, whether the forcing acts on 2D modes alone or a mixture of 2D and 3D modes, more work is needed to further clarify the impact of the forcing dimensionality on this transition.

As the layer height is reduced further, at $H<H_{3D}$, the fraction of energy cascading to large scales increases at the expense of the weakening forward energy cascade. For $H\ll H_{3D}$, scales larger than $H$ show approximately 2D behavior including a forward enstrophy cascade, which is associated with a residual flux of energy to the scale $H$ as well. It was verified in a shell model study\cite{boffetta2011shell} that this residual energy flux at scale $H$ is given by \begin{equation}
D_\nu^\mathcal{E}\propto \eta H^2,\label{eq:residual_forward_flux}
\end{equation} 
where $\eta=\epsilon k_f^2$ is the enstrophy flux, although this remains to be tested in full direct numerical simulations. The energy reaching the scale $H$ is then transported, by 3D interactions, to even smaller scales, \textcolor{black}{where it is dissipated by viscosity}. 

Equation~(\ref{eq:residual_forward_flux}) breaks down when the layer height becomes comparable to the Kolmogorov scale $H\sim \ell_\nu$. There, a third phase of thin-layer turbulence is encountered, where 3D variations are damped out such that $\mathcal{E}_{3D}=0$. In particular, when only the 2D modes are forced, this transition occurs at a critical height $H_{2D}\ll H_{3D}$. For values of $H$ slightly above this threshold, the linear growth rate of 3D modes on the turbulent 2D background flow is random in space and time, leading to rich dynamics, where the energy $\mathcal{E}_{3D}$ of the 3D modes exhibits a scaling
\begin{equation}
    \mathcal{E}_{3D} \propto (H-H_{2D})^{\beta_2},
\end{equation}
with $\beta_2$ observed to be greater than one.\cite{benavides2017critical,alexakis2021symmetry} In fact, in the regime $H \ll \ell_f$, it has been rigorously proven\cite{gallet2015exact,gallet2015exact2,benavides2017critical} that all 3D fluctuations are damped out by viscosity and the flow becomes exactly 2D, such that one recovers the properties described in Sec.~\ref{subsec:2D_turbulence}. 

\begin{figure}
    \centering
     {\large $H>H_{3D}$} \qquad \qquad \qquad \qquad {\large$H<H_{3D}$}\\
    \includegraphics[width=0.23\textwidth]{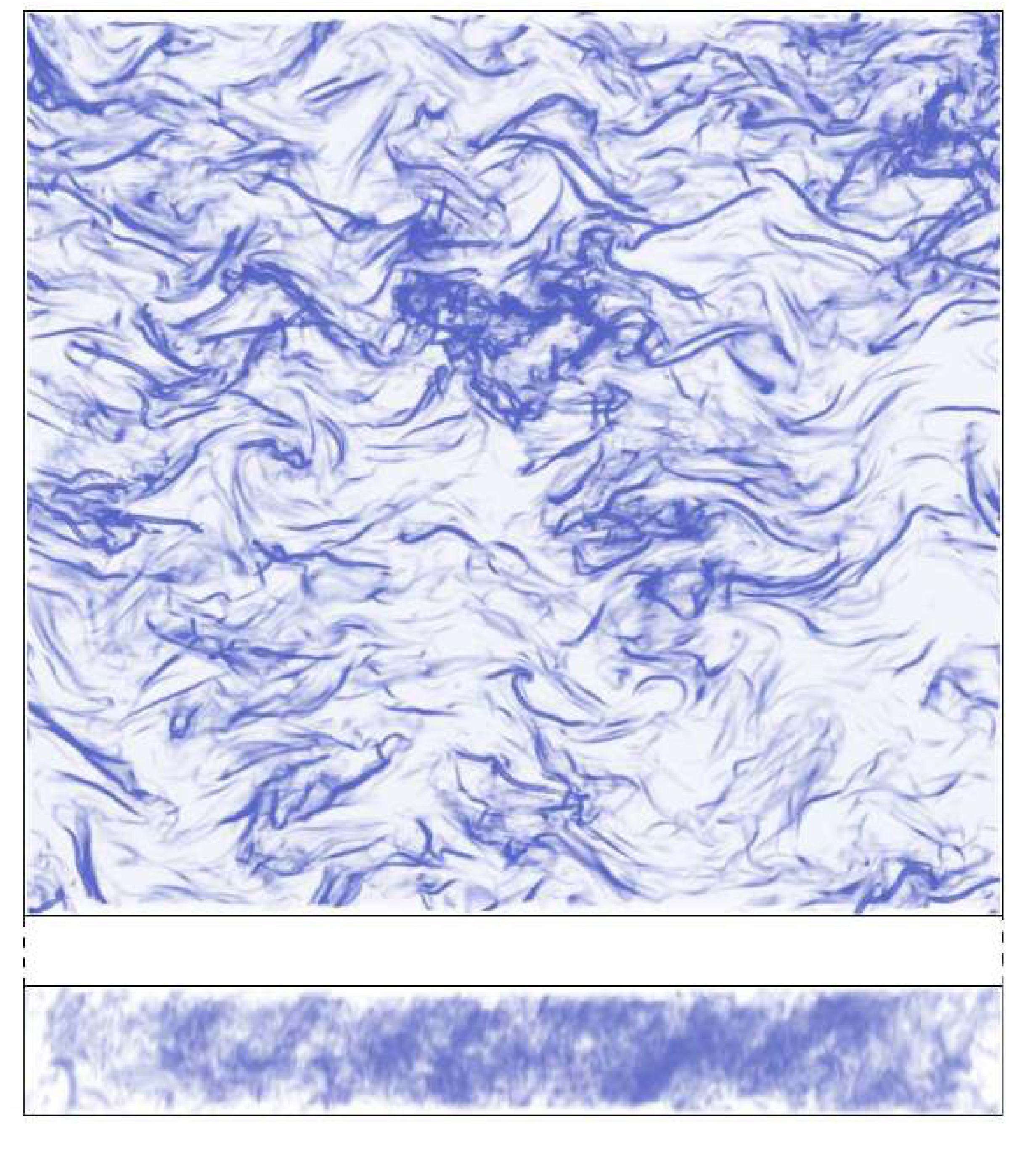}
    \includegraphics[width=0.233\textwidth]{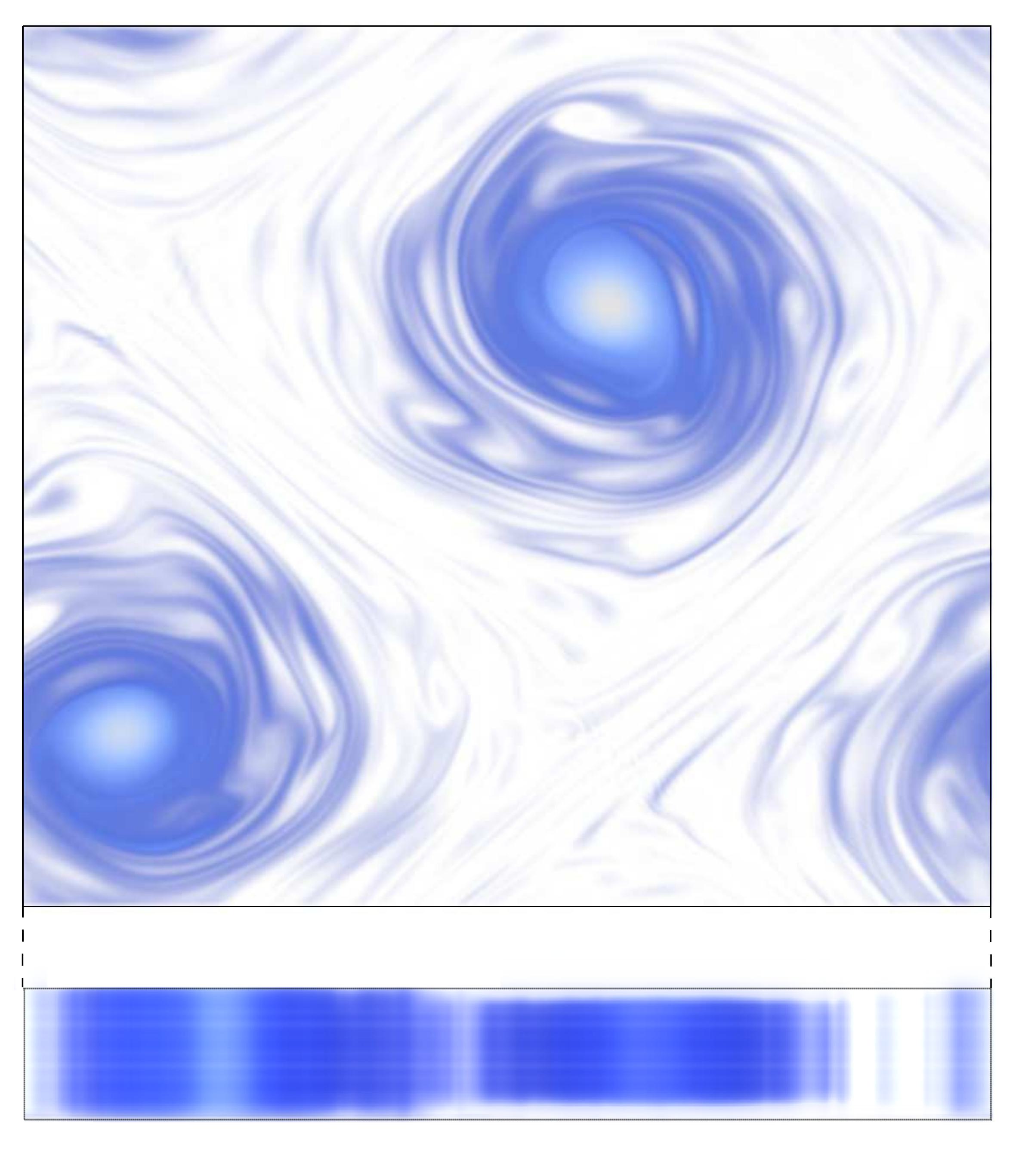}
    \caption{Visualization of $\omega_z^2$ in typical flow states of thin-layer turbulence, showing a top view and a side view of the layer. Left panel: Flow in a deep layer at $H>H_{\rm 3D}$ (see main text). Right panel: large-scale vortex condensate in a thin layer at $H<H_{\rm 3D}$. Figure adapted from van Kan et al. (2019)\cite{van2019condensates}. \textcolor{black}{Source: Adrian van Kan and Alexandros Alexakis, J. Fluid Mech. 864, 490--518, 2019; licensed under a Creative Commons Attribution (CC BY) license.}}
\label{fig:vis_thin_layer_conds}
\end{figure}
\subsubsection{The quasi-2D condensate regime}
Next, we consider the case where the bottom drag is small or absent. In a finite domain, even a small  inverse energy flux then leads to the formation of a large-scale condensate. While numerical simulations of large-scale condensates are computationally very demanding since the saturation of the condensate amplitude is a slow process requiring very long integration times, a few numerical studies exist in the literature\cite{van2019condensates,musacchio2019condensate} which explicitly address condensate formation in thin-layer turbulence. Laboratory experiments, on the other hand, do not suffer from the same restrictions in terms of observation time as simulations, although they are more limited  when it comes to varying the layer geometry or the forcing scale. A number of experimental studies of condensates in thin-layer turbulence have been reported,\cite{xia2009spectrally,shats2010turbulence,xia2011upscale,byrne2011robust,xia2017two,vernet2021turbulence} \textcolor{black}{where the flow is typically driven electromagnetically}, although these have not yet been able to explore the layer height dependence in the same detail.

As in the above discussion of the homogeneous turbulence case, we begin by considering a deep layer of height $H\gg \ell_f$, where a forward energy cascade is present and no condensate forms. As the layer height is reduced to the critical height  $H=H_{3D}$ (discussed above), a weak inverse cascade emerges leading to the formation of a condensate with large-scale energy $\mathcal{E}_{LS}$. It was shown\cite{van2019condensates} (for the case $\alpha=0$) that the transition to the condensate is discontinuous. Specifically, for $H$ slightly larger than $H_{3D}$, no condensate formed when initializing the flow at rest, and the large-scale $\mathcal{E}_{LS}$ was negligible. By contrast, for $H$ just below $H_{3D}$, the large-scale energy $\mathcal{E}_{LS}$ jumped to a finite value, taking up nearly all of the kinetic energy of the flow, indicating the presence of the condensate. Furthermore, the transition was shown to be hysteretic: as $H$ was increased again, the condensate was found to persist at values of $H$ where it did not spontaneously emerge from small-amplitude initial conditions within the integration time. In other words, there is a coexistence of two attractors (illustrated in Fig.~\ref{fig:vis_thin_layer_conds} in terms of $\omega_z^2$) in a range of $H$ near $H_{3D}$. When simulating the system within this parameter range for very long times, one observes rare spontaneous transitions between the two attactors, induced by turbulent fluctuations, occurring at random times. This phenomenon will be described in more detail in section \ref{sec:fluctuation_induced_transitions}. 

When the layer height is reduced further, the large-scale condensate becomes the only stable attractor and its amplitude  in the statistically steady state depends on the dominant dissipation mechanism saturating the inverse cascade. In 2D turbulence with $\alpha=0$, $\mathcal{E}_{LS}\propto \epsilon L^2/\nu$. However, for thin-layer turbulence, it was found\cite{van2019condensates} that for a given layer height $H<H_{3D}$, and sufficiently large Reynolds number, the mechanism for saturation derives from an eddy-viscosity effect due to 3D eddies at scales smaller than $H$. These 3D eddies extract energy from the condensate scales. A flux loop was identified, in which energy injected at the scale $\ell_f$ moves upscale to 
the domain scale $L$ via 2D motions and then back to smaller scales $<H$ via interactions with the smaller-scale 3D eddies. The condensate energy in this case is found\cite{van2019condensates,musacchio2019condensate} not to be inversely proportional to viscosity, but to reach a viscosity-independent value $\mathcal{E}_{LS}\propto f(H/\ell_f) (\epsilon \ell_f)^{2/3}$ instead. A simple three-mode model based on these insights was proposed by van Kan and Alexakis,\cite{van2019condensates} capturing many of these observed features.

As in the homogeneous quasi-2D turbulence case, when the layer height is reduced to the order of the Kolmogorov scale, another phase of turbulence is encountered where 3D perturbations decay viscously. However, the dynamics of 3D modes at $H$ slightly above $H_{2D}$ are strongly impacted by the presence of the condensate. In contrast with homogeneous quasi-2D turbulence, where 3D perturbations can develop anywhere in the flow, the condensate flow is rendered highly inhomogeneous by the presence of large-scale coherent structures, with larger strain and vorticity in some regions than others, while the flow still remains chaotic. The evolution of infinitesimal 3D perturbations on the background of a turbulent 2D condensate was recently investigated  by Seshasayanan et al. in direct numerical simulations,\cite{seshasayanan2020onset,lohani2024effect} revealing that $\mathcal{E}_{3D}$ in this limit 
exhibits a random behavior with long periods of slow viscous decay and short periods of  rapid exponential increase. The periods of increase were found to appear when 
extremes of the vorticity or the strain rate crossed a certain threshold. Considering 2D point vortex flow coupled to localized 3D perturbations, van Kan et al.\cite{van2021intermittency} subsequently showed that the instantaneous growth rate of the energy of the 3D 
perturbations displayed power law distributions that were linked to the power law 
distribution of strain in space. This indicates that the logarithm of the energy of the 3D perturbations undergoes Lévy random motion, explaining the sudden jumps in the growth 
of the perturbation energy observed in the simulations of Seshasayanan et al. This led to the investigation of this type of random process in simplified stochastic dynamical systems, where a new type of intermittency was identified (distinct from the dissipative intermittency of 3D turbulence mentioned in Sec.~\ref{sec:bg}), so-called Lévy on-off intermittency,\cite{van2021levy,van2023noise} \textcolor{black}{namely, stochastic dynamics near a phase transition subject to multiplicative power-law noise.} In particular, this theory predicts a distinct power-law behavior of the 3D energy
\begin{equation}
    \mathcal{E}_{3D} \propto (H-H_{2D})^{\beta_3},
\end{equation}
where the exponent $\beta_3$ depends on the noise parameters.\cite{van2021levy,van2023noise} This behavior suggested by the point-vortex model is 
yet to be confirmed explicitly in direct numerical simulations.\\ 

Thin-layer turbulence is arguably the simplest, but by no means the only system to display dimensional transitions. Below, we summarize similar phenomena in a selection of more complex systems, describing how additional physical effects alter the transition phenomenology.

\subsection{Rapidly rotating turbulence}
\begin{figure}
    \centering
    \includegraphics[width=\linewidth]{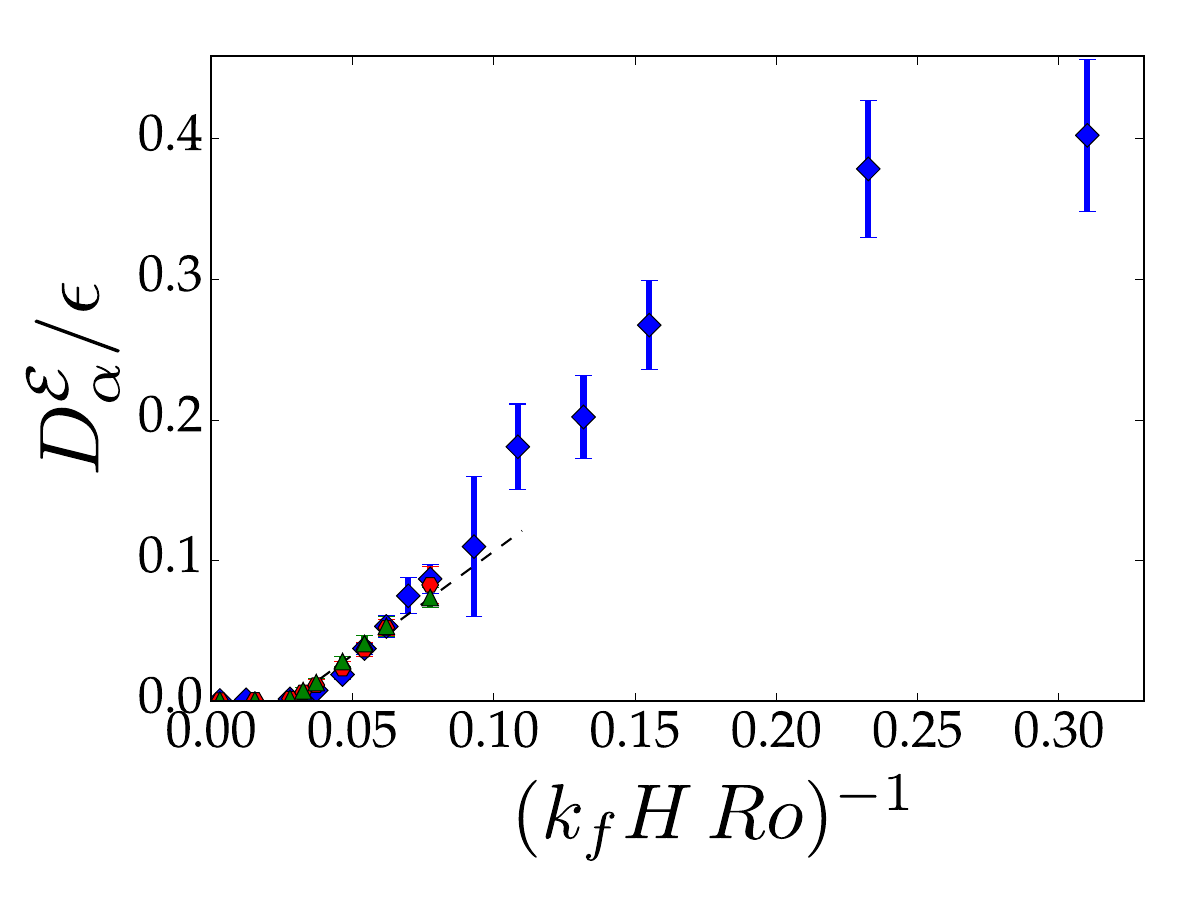}
    \caption{\textcolor{black}{Fraction of energy cascading upscale in steady state, $D_\alpha^\mathcal{E}$ as a function of the parameter $(k_fH\, Ro)^{-1}$, showing a transition from zero to nonzero values at a threshold value around $(k_fH\, Ro)^{-1}\approx 0.03$, with an approximately linear scaling near onset. Different symbols indicate different forcing scales and hyperviscous Reynolds numbers. Reproduced with permission from  J. Fluid Mech., 899, A33 (2020). Copyright 2020 Cambridge University Press.}}
    \label{fig:phase_transition}
\end{figure}

Flows in a reference frame rotating at some rate $\Omega$ are impacted by the Coriolis force, which tends to suppress variations in velocity along the axis of rotation, a result known as the Taylor-Proudman theorem.\cite{proudman1916motion,taylor1917motion} Therefore, rapidly rotating turbulence is quasi-2D and displays split energy cascades. The strength of rotational effects is quantified by the Rossby number ${\rm Ro} = \epsilon^{1/3}/(\Omega \ell_f^{2/3})$, with small $\rm Ro$ corresponding to rapid rotation. While the effect of rotation was already included in the early work of Smith et al.,\cite{smith1996crossover} detailed investigations of the transition from 3D turbulence to quasi-2D turbulence were carried out subsequently by Deusebio et al.\cite{deusebio2014dimensional} and Pestana and Hickel.\cite{pestana2019regime} Whereas for slowly rotating flows at $\rm Ro \gtrsim 1$ one finds $H_{3D}\propto \ell_f$, this scaling is modified in the rapidly rotating regime ($\rm Ro \ll 1$) to become 
\begin{equation}
    H_{3D}\propto \ell_f/{\rm Ro},
\end{equation}
which was also verified in the limit $\rm Ro \to 0$ using a reduced, asymptotically valid set of equations.\cite{van2020critical} \textcolor{black}{Figure \ref{fig:phase_transition} shows the fraction of kinetic energy cascading upscale as a function of $(k_f H\, Ro)^{-1}$ transitions from zero to nonzero values at a critical threshold $(k_f H\, {\rm Ro})^{-1}\approx 0.03$, with approximately linear scaling near onset (different symbols indicate different Reynolds numbers and forcing scales). }

The dimensional transition in rotating turbulence is also characterized by a power-law scaling of the inverse energy flux with the distance from a critical value of $\rm Ro$, not unlike the behavior seen in thin-layer turbulence as a function of $H_{3D}-H$. However, despite this similarity, there are important differences between the two cases. In particular, at intermediate values of $\rm Ro$, it was found by Clark Di Leoni et al.\cite{di2020phase} that the inverse cascade is arrested by the formation a long-lived, regular array (or crystal) of cyclones, while anticyclones decayed. Such structures are an alternative organized flow state in addition to large-scale condensates, and are widely observed, from experiments of rotating convection \cite{boubnov1986experimental} to magnetized electron columns \cite{fine1995relaxation} and active matter.\cite{riedel2005self,xu2024self} Another particularly striking example of a vortex crystal was observed at the North pole of Jupiter by the Juno satellite mission.\cite{adriani2018clusters} The results of Clark Di Leoni et al. could provide an additional explanation for this phenomenon, complementing interpretations based on barotropic quasi-geostrophic flow on the sphere.\cite{siegelman2022polar}

Large-scale condensates in rotating turbulence were also investigated.\cite{alexakis2015rotating,seshasayanan2018condensates,yokoyama2017hysteretic} In particular, bistability between the condensate state and 3D turbulence close to $H_{3D}$ has also been described in that case. However, rotating turbulence is more complex, in particular due to the presence of inertial waves, whose restoring force is the Coriolis force, and whose frequency increases with $1/{\rm Ro}$, posing numerical challenges. This added complexity has thus far prevented a more detailed investigation. Fortunately,  however, advanced experimental platforms have been able to probe the formation of large-scale flow structures directly in the laboratory, confirming the presence of bidirectional cascades, and allowing one to study the role of nonlinearly interacting inertial waves (also called inertial wave turbulence) in the dynamics.\cite{lamriben2011direct,campagne2014direct,gallet2014scale,yarom2014experimental,kolvin2009energy,brunet2020shortcut,monsalve2020quantitative} These and other, forthcoming laboratory experiments are well suited for studying the long-time evolution of rotating 
turbulence, including close to critical points. Future work is expected to confront numerical results with
experimental observations, as well as uncovering new physics.

\subsection{(Stably) Stratified turbulence}
Stably stratified turbulence, or stratified turbulence for short, refers to turbulence under the influence of gravity $-g \hat{\mathbf{z}}$ and an imposed, stable background density gradient $S=-\rho_0^{-1} {\rm d} \rho(z) / {\rm d}z>0$, where $\rho_0$ is a reference density. Like rotating flow, stratified flow supports wave motion, namely internal waves whose frequency depends on stratification. Unlike rotation, however, stratification acts to suppress the formation of 2D turbulence: by inhibiting vertical motion due to the energy cost associated with lifting up heavy fluid parcels, it leads to the formation of thin layers, creating strong vertical gradients in the process. The strength of stratification is measured by the nondimensional Froude number ${\rm Fr} = \epsilon^{1/3}/(gS)^{1/2}/\ell_f^{2/3} $, where small values of $\rm Fr$ indicate strong stratification. Sozza et al.\cite{sozza2015dimensional} showed that the critical height, at which the bidirectional energy cascade emerges, decreases with increasing stratification strength as
\begin{equation}
    H_{3D} \propto \ell_f {\rm Fr}.
\end{equation}
This indicates that in strongly stratified flows, inverse energy cascades can appear at much smaller $H$ than in the absence of stratification. A more detailed presentation can be found in the recent review by Boffetta.\cite{boffetta2023dimensional} More work will be needed to further characterize the physics of the phase transitions in stratified turbulence and the properties of large-scale condensates in this system, which have not been investigated to date.
\subsection{Rotating stratified turbulence}
\begin{figure}
    \centering
    \includegraphics[width=0.7\linewidth]{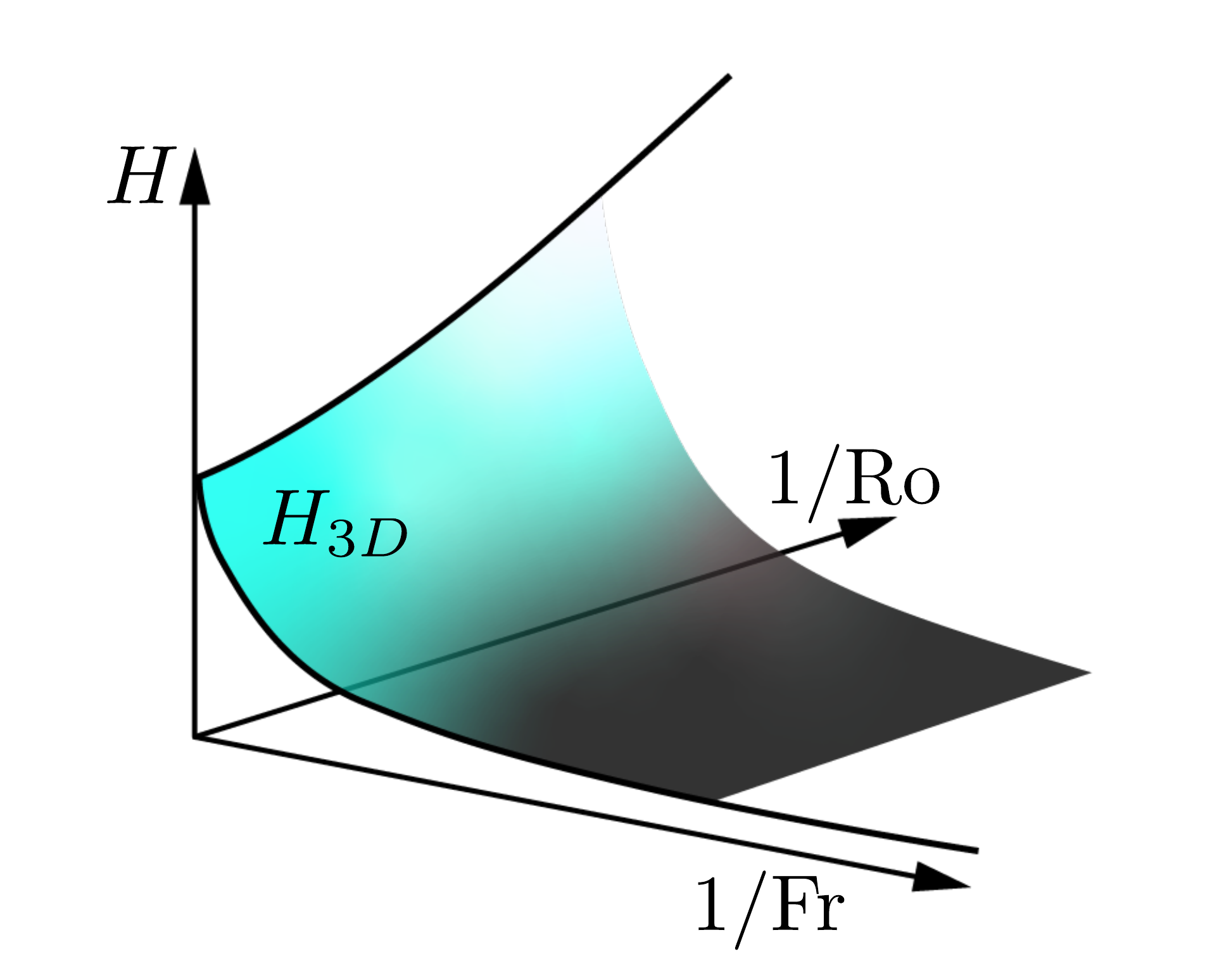}
    \caption{Sketch of the phase space of rotating stratified turbulence spanned by the layer height $H$, the inverse Rossby number $1/\rm Ro$ and the inverse Froude number $1/\rm Fr$. A hypothetical critical surface is indicated (based on the results discussed in the text), representing the height $H_{3D}$ where the bidirectional energy cascade emerges. Above the surface (at $H>H_{3D}$), only a forward energy cascade is present, while below the surface ($H<H_{3D}$) there is a bidirectional cascade. }
    \label{fig:sketch_parameter_space_rot_strat_turb}
\end{figure}

Rotating stratified turbulence can be regarded as the simplest model of the dynamics within a dry planetary atmosphere, and numerous studies have been devoted to studying its cascade properties.\cite{bartello1995geostrophic,marino2013inverse,pouquet2017dual} However, even though it is highly idealized, rotating stratified turbulence continues to challenge our understanding. For instance, it has only very recently become possible in direct numerical simulations to achieve values of $\rm Ro/Fr$ that are realistic for (part of) Earth's atmosphere, as described in the work of Alexakis et al. (2024).\cite{alexakis2024large} Rotating stratified turbulence is special since it combines the competing effects of two-dimensionalization by rotation and the tendency for layering of stratification. Moreover, it features gravito-inertial waves, which add to the complexity of the dynamics. The result is a rich 3D parameter space (of which a sketch is shown in Fig.~\ref{fig:sketch_parameter_space_rot_strat_turb}), spanned by the nondimensional layer height $H/\ell_f$, the Rossby number $\rm Ro$ and the Froude number $\rm Fr$, which remains incompletely explored. From a practical perspective, it is challenging to establish the existence of a critical point or a critical surface within a 3D parameter space, since sampling any one set of parameters requires an expensive numerical simulation. 

Reducing the parameter space by using asymptotic methods to obtain simplified, leading-order equations in the regime $\rm Ro \ll 1 $, $H=O(1/{\rm Ro})$, $\rm Fr =O(1)$,  van Kan and Alexakis (2022)\cite{van2022energy} showed that the parameter space in that limit is subdivided into at least three different 
phases, one with no inverse cascade, one rotation-dominated regime with an inverse cascade due to a two-dimensionalization and a third strongly stratified regime with an inverse cascade due to quasi-geostrophic dynamics. This indicates that the phase space of rotating stratified turbulence is highly complex, more so than  the sketch in Fig.~\ref{fig:sketch_parameter_space_rot_strat_turb} can convey, and a  further meticulous exploration of this complexity is needed, including in suitable asymptotic limits. 
\subsection{Convection}
Convection refers to fluid motion induced by buoyancy in the presence of an imposed unstable background density gradient ($S<0$). The paradigmatic model of this process is Rayleigh-Bénard convection, where the fluid is contained between a hot bottom boundary and a cold top boundary\textcolor{black}{, both of which are typically assumed to be either of no-slip or stress-free type}. Convection induces three-dimensional vertical motion. The resulting dynamics are highly nontrivial, in particular since the buoyant forcing is self-organized in response to the flow, rather than being an externally imposed body force. Despite these differences, it has been found that convection, under highly anisotropic conditions, can also display quasi-2D behavior and the formation of large-scale flow structures.

The most well-studied example is that of rapidly rotating convection.\cite{julien2012statistical,ecke2023turbulent} There, for sufficiently small Rossby numbers, an inverse energy cascade forms and results in the formation of large-scale vortices or jets,\cite{guervilly2014large,favier2014inverse,rubio2014upscale,guervilly2017jets,julien2018impact,maffei2021inverse} not unlike the systems discussed in the previous sections.  Furthermore, the behavior of the large-scale vortex 
condensates also resembles thin-layer and rotating turbulence, since they emerge via a discontinuous transition.\cite{favier2019subcritical,de2022discontinuous} However, the case of convection has additional complexity, in particular due to the presence of thermal and momentum boundary layers. \textcolor{black}{It is known that the emergence of the inverse cascade is impacted by the boundary conditions. In the presence of stress-free boundaries, at moderately rapid rotation, i.e, moderately small Rossby number and Ekman number $Ek=\sqrt{\nu/(2\Omega H^2)}$, large-scale vortices form, which are suppressed with no-slip boundaries by viscous Ekman boundary layers\cite{kunnen2016transition}, where Ekman pumping plays an important role.\cite{julien2016nonlinear} Nonetheless, at sufficiently rapid rotation rates (sufficiently low $Ek$) the inverse energy cascade and associated large-scale vortices are also observed in the case of no-slip boundaries.\cite{song2024direct} Importantly, the corresponding problem of the robustness of inverse cascades and condensation with respect to no-slip boundaries, which are typically present in laboratory experiments, remains to be studied in mechanically forced thin-layer turbulence.}
Recent progress in numerically simulating convection at very low Rossby numbers \textcolor{black}{and Ekman numbers as low as $Ek=10^{-15}$} has furthermore revealed a phase transition between fully rotationally constrained geostrophic turbulence and a regime where \textcolor{black}{(thermal)} boundary layers exhibit ageostrophic turbulence,\cite{van2024bridging,julien2024rescaled} which had been predicted theoretically.\cite{julien2012heat} 

Another interesting case where large-scale structures can form in the absence of rotation is that of small-aspect-ratio domains where a constant heat flux is applied to the layer,\cite{vieweg2021supergranule,vieweg2024supergranule} rather than the more classical setup of constant temperatures at the boundaries.

\section{Multistability and Fluctuation-induced transitions \label{sec:fluctuation_induced_transitions}}
 As was mentioned in the previous section, turbulent flows can display multistability between distinct coexisting attractors. Indeed, this behavior is commonly observed across different physical systems. In many cases, these attractors are metastable, in the sense that they have a finite lifetime due to turbulent fluctuations which can induce spontaneous transitions between the different stable states. Such transitions can have different manifestations in terms of the flow morphologies involved, including reversals of large-scale mean fields, transitions between small-scale turbulence and large-scale coherent structures, or transitions between qualitatively distinct large-scale flows. New discoveries have recently been made regarding these different forms of multistability and associated fluctuation-induced transitions in turbulence, which we discuss below.

\subsection{Reversals of mean fields}
Spontaneously reversing large-scale fields are commonly found in turbulent fluid flows. An important example is the Earth's magnetic field, which is thought to arise from the dynamo effect \cite{fauve2003dynamo} associated with motions in the liquid metal outer core, and whose polarity is known to reverse at random times.\cite{jacobs1994reversals} By contrast, the Sun's magnetic field follows the more regular 22-year solar cycle.\cite{charbonneau2014solar} Magnetohydrodynamic dynamo reversals have also been reproduced in laboratory experiments,\cite{berhanu2007magnetic} and were analyzed extensively at the theoretical level.\cite{petrelis2009simple,gallet2012reversals} An additional example of a reversing flow is found at high altitudes in the Earth's atmosphere, where the so-called quasi-biennial oscillation consists in reversals of the mean zonal wind approximately every two years,\cite{baldwin2001quasi} a phenomenon that has also been studied using laboratory analogs.\cite{plumb1978instability,semin2018nonlinear}
In addition, reversals are also common in convectively driven flows and have been studied in detail in that context.\cite{sugiyama2010flow,ni2015reversals,wang2018mechanism,chen2019emergence,winchester2021zonal,liu2022staircase,liu2023fixed} 

Another setting where random reversals are observed is that of confined, large-scale, 2D or quasi-2D turbulent flows, which have been studied both in the laboratory\cite{sommeria1986experimental,michel2016bifurcations,fauve2017instabilities,pereira20191} and in theoretical work.\cite{mishra2015dynamics,shukla2016statistical,dallas2020transitions} Shukla et al.\cite{shukla2016statistical} introduced the approach of investigating the statistics of the large-scale modes based on the truncated Euler equation of ideal fluids, retaining only a finite number $N$ of Fourier modes with wave numbers below a chosen cutoff. Such truncations naturally arise in any standard pseudospectral PDE solver.  They considered the flow inside a square domain $[0,\pi]^2$ with free-slip boundary conditions, where the stream function can be expanded as $\psi(x,y) = \sum_{n,m} \hat{\psi}_{n,m}\sin(nx)\sin(my)$, with the sum being over a truncated set containing $N$ modes. Shukla et al. compared direct numerical simulations of the 2D Navier-Stokes equation for a viscous fluid with corresponding solutions of the 2D truncated Euler equation \textcolor{black}{in the condensate regime} and were able show good qualitative \textcolor{black}{(but not quantitative)} agreement between the statistics \textcolor{black}{of the largest-scale mode in the two cases, with the appearance of reversals of the large-scale circulation, which become less and less frequent as friction is reduced, leading to a stronger condensate.Similar findings were also made in a forced 2D shear flow.\cite{dallas2020transitions_prf}} 

\begin{figure}
    \centering
    \includegraphics[width=0.5\textwidth]{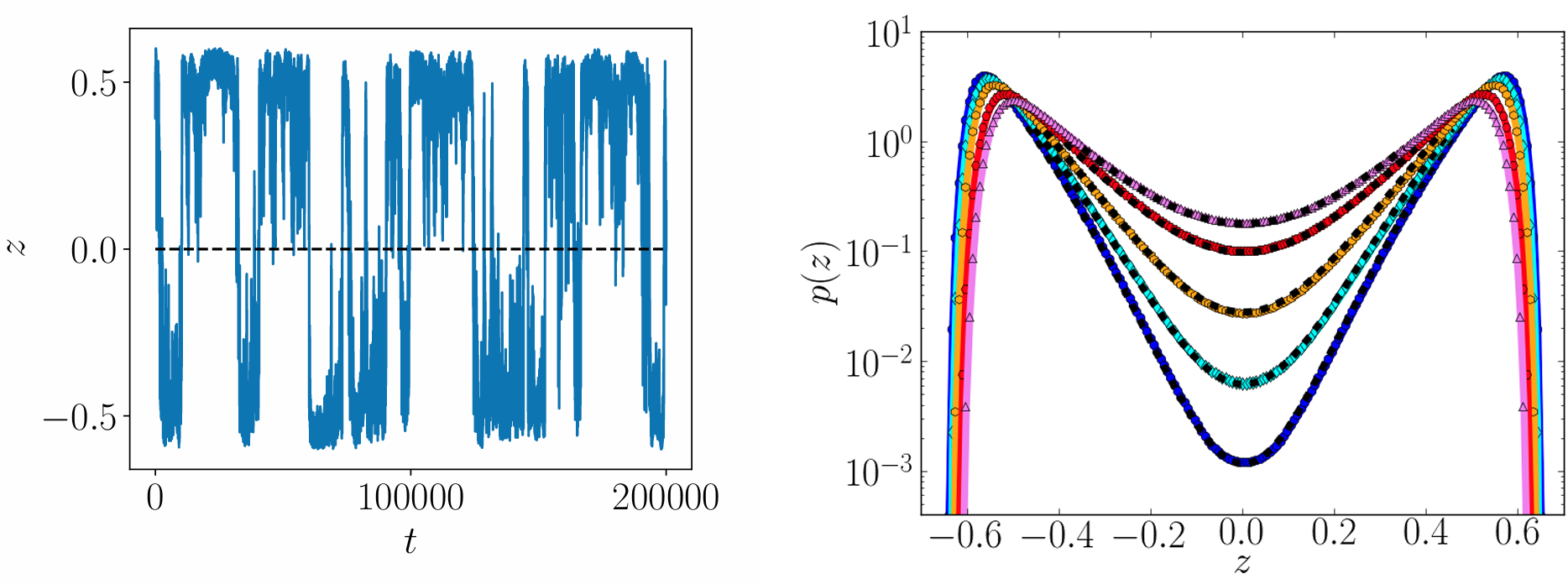}
    \caption{Left panel: Spontaneous reversals seen in the time series of the large-scale mode $\hat{\psi}_{1,1}$ in a confined 2D flow described by the Euler equation truncated to $N=13$ modes, obtained from numerical simulations. Right panel: probability density function (PDF) $p(\hat{\psi}_{1,1})$ versus large-scale circulation $\hat{\psi}_{1,1}$ with different colored lines representing different values of the control parameter $\delta = \Omega/(k_2^2 \mathcal{E}) - 1$ ($\delta>0$ increases from bottom to top). The black dashed lines indicate the theoretically predicted form of $p(\hat{\psi}_{1,1})$ at different values of $\delta$, which agrees well with the numerical result. \textcolor{black}{Reproduced with permission from Phil. Trans. R. Soc. A. 380:20210049 (2022). Copyright 2022 Royal Society.}  }
    \label{fig:ls_mode_reversals_tee}
\end{figure}

Recently, van Kan et al.\cite{van2022geometric} went beyond the work of Shukla et al., using a geometric approach to explicitly compute $N$-dimensional phase space integrals to derive exact statistical results for this problem in the framework of microcanonical statistical mechanics based on the simultaneous conservation of energy $\mathcal{E}$ and enstrophy $\Omega$ in the truncated Euler equation. They showed that the onset of reversals (sign changes in the large-scale mode $\hat{\psi}_{1,1}$ associated with wave number $k_1=\sqrt{2}$) occurs when the control parameter $\delta \equiv \Omega/(k_2^2 \mathcal{E}) - 1$ passes through zero, with $k_2=2$ being the second smallest wave number in the system. Specifically, for $\delta<0$, no reversals are possible in this system due to ergodicity breaking, but for $\delta>0$, there are chaotic reversals of the large-scale flow, see Fig.~\ref{fig:ls_mode_reversals_tee}(a). The authors analytically derived the form of the probability distribution of the large-scale mode in the problem near the onset of reversals ($0<\delta \ll 1$),
\begin{equation}
    p(\hat{\psi}_{1,1}) \propto \left\lbrace(k_2^2-k_1^2)\hat{\psi}_{1,1}^2 + k_2^2\delta\, \mathcal{E}\right\rbrace^\frac{N-5}{2}, 
\end{equation}
 which was validated in DNS as shown in Fig.~\ref{fig:ls_mode_reversals_tee}(b). They further deduced a power-law relation between the mean first-passage time and $\delta$ near onset. These results, based on the truncated Euler equation, are complementary to existing works in the Robert-Sommeria-Miller framework. A similar approach based on Galerkin truncation could be fruitful for the 3D Euler equation, where energy and helicity are conserved instead of energy and enstrophy, or even more complex flows with other ideal invariants, potentially opening the door to analytical predictions in these systems as well.

\subsection{Transitions between flows with and without large-scale structure}
Certain fluid systems can sustain small-scale turbulence and large-scale organized flow at the same external parameters, depending on initial conditions. In quasi-2D turbulence, this occurs near $H=H_{3D}$, or more generally near the onset of the inverse cascade. We reiterate that this type of bistability has been observed in body-forced thin-layer turbulence,\cite{van2019rare,de2022bistability} as well as rotating turbulence,\cite{yokoyama2017hysteretic} where large-scale vortices and small-scale 3D turbulence coexist over a range of the control parameter (layer height/rotation rate), and a  similar type of subcritical transition was identified in rotating convection.\cite{favier2019subcritical,de2022discontinuous} Spontaneous transitions have not been observed in all cases, since they may be exceedingly rare, depending on physical control parameters, which requires very long integration times that pose a numerical challenge.

 In the specific case of thin-layer turbulence, it was shown that the lifetimes of the large-scale vortex and 3D turbulence states, respectively, change faster than exponentially with $H$,\cite{van2019rare} and may possibly become infinite at a finite threshold height.\cite{de2022bistability} Such a superexponential relationship between typical lifetimes of metastable turbulent states and a control parameter is reminiscent of the (physically very different) problem of the transition to turbulence in pipe flow, where turbulent puffs decay or split on time scales depending on a double exponential of the Reynolds number $\rm Re$,\cite{avila2023transition} with equality between these timescales determining the critical value of $\rm Re$ for the onset of turbulence. The critical height of the layer for the onset of large-scale condensation in thin-layer turbulence, in the sense of statistical preference of one attractor over the other, may be defined in a similar way.

 While experiments have examined the dynamics of large-scale vortices within thin-layer flows in the laboratory,\cite{xia2011upscale} more work is needed to ascertain the robustness of the numerical results with respect to realistic experimental boundary conditions. In particular, fluctuation-induced transitions between large-scale and small-scale turbulence at a given set of control parameters remain to be studied in the laboratory. It is encouraging in this regard that recent experiments on shallow layers of quantum fluids report related behavior.\cite{novotny2024critical}

\subsection{Transitions between distinct large-scale flows}
A third important class of fluctuation-induced transitions in turbulent flows involves the competition between distinct large-scale flow structures, such as large-scale vortices and jets, i.e., bands of unidirectional winds, typically forming in the East-West (zonal) direction as a consequence of planetary rotation. Large-scale vortices and zonal jets are widely observed planetary atmospheres including on Earth, and feature prominently on Jupiter. 

The number of zonal jets in a planetary atmosphere can change over time. On the time scale of the observational record, the Jovian zonal (east-west) jets have been remarkably unchanged in number and intensity,\cite{ingersoll2004dynamics} despite turbulent dynamics. However, various idealized models of geophysical and astrophysical relevance allow us to study the dynamics far beyond the observational record, revealing the possibility of rare transitions between different numbers of jets induced by turbulent fluctuations. This has been found in different models, including barotropic beta plane turbulence (where the latitudinal variation in the Coriolis force is taken into account)\cite{rhines1975waves,bouchet2019rare,cope2020dynamics,simonnet2021multistability} and rotating thermal convection.\cite{guervilly2017jets} Bistability between atmospheric flow states with and without a strong equatorial eastward jet (superrotation) has furthermore been found in climate models,\cite{herbert2020atmospheric} indicating a possible scenario of abrupt climate change.

Earth's mid-latitude jet stream can spontaneously develop stationary meanders, which is referred to as a blocking event,\cite{nakamura2018atmospheric} typically associated with extreme heat waves in Europe. Transitions of this type have been reproduced in simple laboratory analogs.\cite{weeks1997transitions} In the oceanic context, the paths of Western boundary currents, such as the Kuroshio current, display a similar bimodality across different years.\cite{schmeits2001bimodal} The oceanic thermohaline circulation provides another important instance of bistability and noise-induced transitions between on and off states of the Gulf Stream.\cite{timmermann2000noise} More generally, multistability is common in the climate system as a whole.\cite{margazoglou2021dynamical,ragon2022robustness}

\begin{figure}
    \centering
    \includegraphics[width=\linewidth]{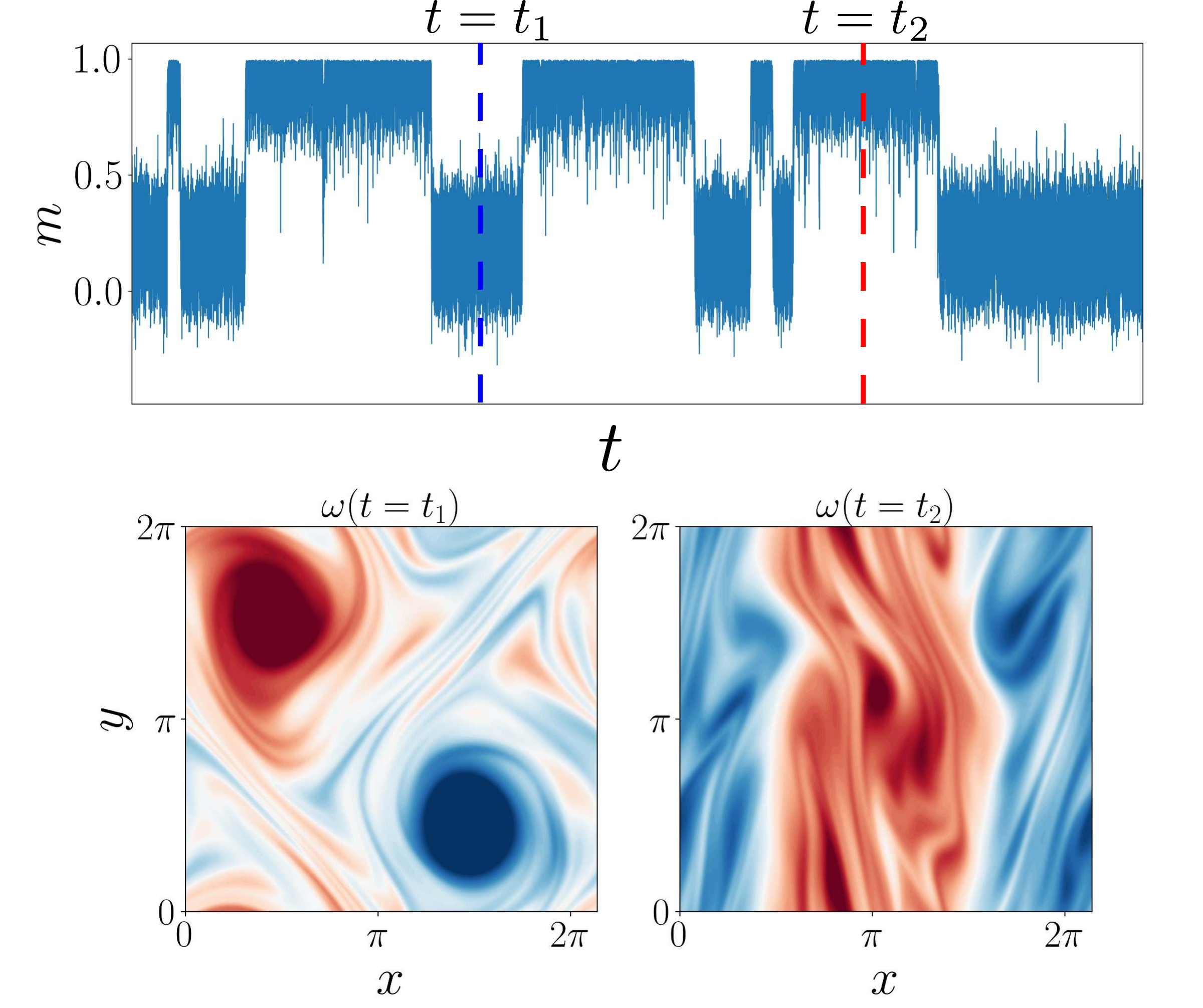}
    \caption{Rare, fluctuation-induced transitions between `hurricane-like' large-scale vortices and unidirectional jets (not unlike the Earth's jet stream) in 2D turbulence within a weakly elongated, periodic domain (at $\Delta=1.06$, i.e. $6\%$ elongation in the $x$-direction). \textcolor{black}{Top: time series of polarity $m=\langle v^2 - u^2\rangle / \langle u^2+v^2\rangle$ of the 2D velocity field $\mathbf{u}=(u,v)$, where $\langle \cdot \rangle$ represents the domain average, showing transitions from $m$ close to zero (large-scale vortices) to $m\approx 1 = 1$ (unidirectional jets in the $y$ direction).} Bottom: snapshots of $\omega_z$ in large-scale vortex (left) and jet (right) \textcolor{black}{representative of states at the times $t=t_1,t_2$ indicated in the top panel by vertical dashed lines. Figure reproduced from Phys. Rev. Fluids 9.6 (2024): 064605. Copyright 2024 American Physical Society.}}
    \label{fig:illustr_jets_vortices}
\end{figure}

In a simpler setting, condensate flows can also spontaneously transition between large-scale vortices and jets (illustrated in Fig.~\ref{fig:illustr_jets_vortices}). This was first described in the context of stochastically forced 2D turbulence (with $\alpha=0$) within rectangular (nearly square) domains of aspect ratio $\Delta=L_x/L_y$ by Bouchet and Simonnet,\cite{bouchet2009random} who emphasized a statistical-mechanics-based interpretation of this phenomenon. The jets described by Bouchet and Simonnet are oriented parallel to the short side of the rectangular domain.

In a recent study, Xu et al.\cite{xu2023fluctuation} went significantly beyond the work of Bouchet and Simonnet, with extensive DNS of the same setup and substantially longer integration times in excess of $10^5$ viscous diffusion times ($200$ times longer than previously available simulations), providing a detailed statistical analysis of the lifetimes of large-scale vortices and jets in the bistable range of aspect ratios. This revealed an approximately exponential dependence of the mean lifetimes of large-scale vortices and jets on the deviation $\Delta-1$ of the aspect ratio from unity: the mean lifetime of large-scale vortices decreases as the domain is elongated from the square shape, while the lifetime of jets increases in the process. The exponential dependence resembles the classical Eyring-Kramers formula \cite{mel1991kramers} for the escape time of an overdamped particle from a potential well due to thermal fluctuations, while differing from the faster-than-exponential dependence found in thin-layer turbulence.  

An important role was found to be played by the kinetic energy gap formed between large-scale vortices and jets, which exists due to the scale-dependence of purely viscous damping ($\alpha=0$): jets vary in the elongated dimension and therefore have a larger scale, experiencing weaker viscous damping than the large-scale vortices and thus saturating at higher amplitude. Transitions between the two states were found to be rarer for bigger energy gaps. Xu et al. also computed the phase space trajectories of the system during transitions between the two attractors, showing that the path starting from the jet and ending in large-scale vortex state differs from the path starting in the large-scale vortex and ending in the jet state, a finding which is in line with previous results for jets on a beta plane.\cite{bouchet2019rare} Rare transitions between different numbers of jets were also observed at larger values of $\Delta\gtrsim2.5$.

The wealth of examples described above shows that multistability and noise-induced transitions are ubiquitous in turbulence, but further efforts numerical and theoretical will be needed to better understand the complex dynamics associated with them. For example, it is yet unknown how the properties of fluctuation-induced transitions between jets and vortices described above are impacted by 3D modes in a quasi-2D setting.

\section{Other transitions in two-dimensional turbulence\label{sec:other_transitions_2D_turbulence}}
Phase transitions have also been identified in different variants of 2D turbulent flows. Here, we focus on the examples of 2D turbulence over topography and instability-driven turbulence, summarizing recent progress on these problems.
\subsection{Turbulence over topography}
\textcolor{black}{The following section is largely inspired by the work of Gallet\cite{gallet2024two}, with the permission of the author.}
 \textcolor{black}{Highly anisotropic fluid flows} above topography provide an idealized model for large-scale geophysical flows and also represent a proving ground for studying how spatial inhomogeneity impacts large-scale structure formation in \textcolor{black}{(quasi-)two-dimensional} flows. Interesting oceanic examples include the Antarctic Circumpolar Current, where turbulent eddies are funneled by and may be locked to bottom topography \cite{straub1993transport} and, at smaller scales, the Lofoten basin vortex,\cite{soiland2013structure} a highly persistent anticyclone located at the center of the basin.

 The study of \textcolor{black}{quasi-geostrophic flows, in which the horizontal momentum balance is dominated by the Coriolis force associated with planetary rotation balanced by pressure gradients and advection is primarily horizontal}, over bottom topography has long been was interlinked with the broader study of large-scale self-organization in highly anisotropic flows. Bretherton and Haidvogel\cite{bretherton1976two} applied the concept of `selective decay' (an organizational principle borrowed from plasma physics) to such flows, which states that forward-cascading invariants should be minimized for a given value of the inverse-cascading invariants. Based on the inverse energy cascade and forward enstrophy cascade of 2D turbulence \textcolor{black}{(both of which also arise in quasi-geostrophic turbulence)}, Bretherton and Haidvogel thus predicted that the system evolves towards a state characterized by a minimum in the enstrophy, while conserving its initial energy. The limitation of this and similar results, as in homogeneous 2D turbulence, is that they are formulated based on exact energy conservation, and may not apply in the presence of forcing or dissipation.

  Siegelman and Young\cite{siegelman2023two} recently reconsidered the problem of unforced and weakly damped quasi-geostrophic flow over random rough topography by means of numerical simulations. As one would expect from selective decay, enstrophy decreased over time in their simulations, while energy was approximately constant. However, at late times, the flows observed by Siegelman and Young departed from the predictions of selective decay. Specifically, they identified striking differences between flows with low kinetic energy $\mathcal{E}$ and flows with high $\mathcal{E}$, separated by a threshold energy $\mathcal{E}_{\rm t}$ associated with a given topography: flows with $\mathcal{E}<\mathcal{E}_{\rm t}$ feature (weak) vortices which are strongly correlated with the bottom topography, with anticyclones locked to topographic depressions and cyclones to topographic elevations, a phenomenon not predicted by selective decay. At $\mathcal{E}> \mathcal{E}_{\rm t}$, strong vortices are able to detach themselves from the topography and the vorticity-topography correlation disappears. 
 
\begin{figure}
 \centering
    (a) \\
    \includegraphics[width=0.7\linewidth]{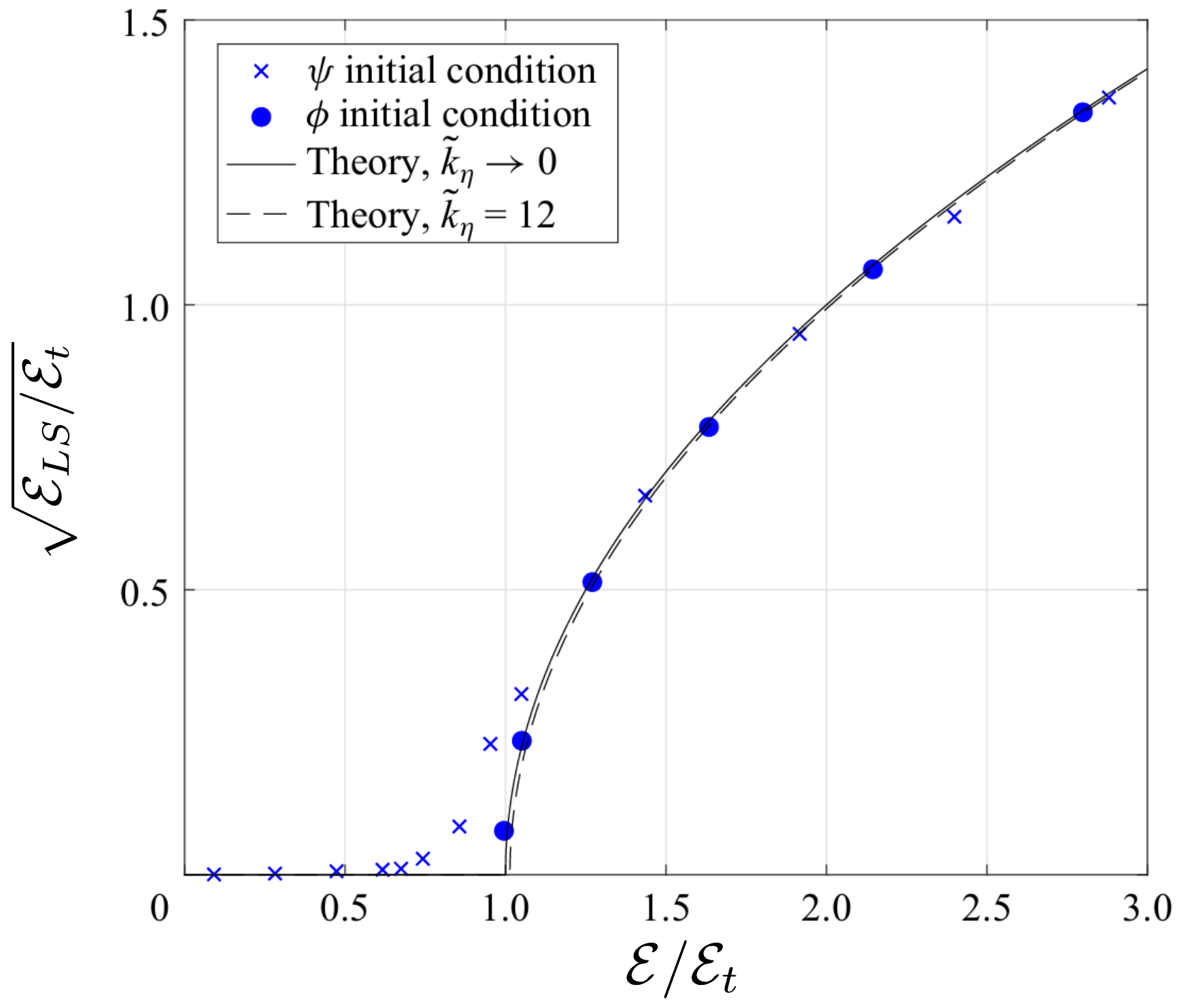}\\
    \centering
    (b)  \qquad\qquad\qquad\qquad (c) \qquad\qquad\qquad\qquad (d) \\    \includegraphics[width=0.31\linewidth]{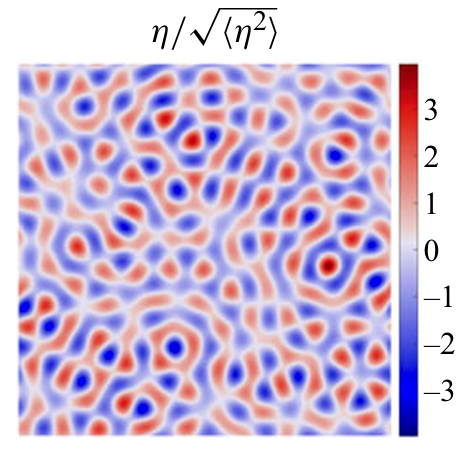}
    \includegraphics[width=0.33\linewidth]{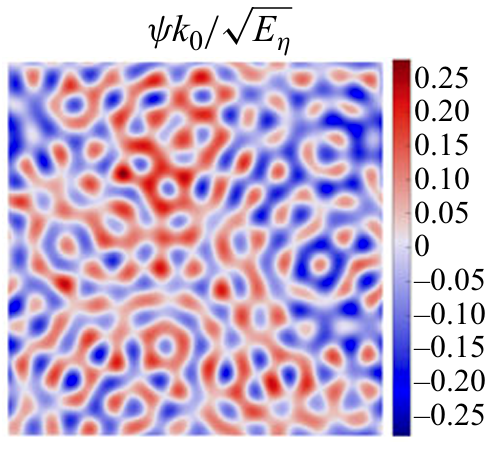}
    \includegraphics[width=0.33\linewidth]{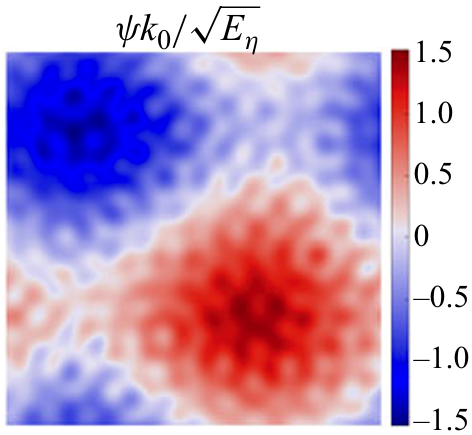}
        Topography \quad  \quad Streamfunction \quad \quad Streamfunction \\
        \hspace{2cm} ($\mathcal{E}<\mathcal{E}_{\rm t}$) \hspace{1.5cm} ($\mathcal{E}>\mathcal{E}_{\rm t}$)
    \caption{\textcolor{black}{Panel (a): Nondimensional condensate amplitude versus total energy revealing phase transition in large-scale structure, in agreement with generalized selective decay, with only a weak dependence on initial conditions.} Panel (b): example of single-scale topography over which 2D turbulence was studied in.\cite{gallet2024two} Panel (c): snapshot of streamfunction for low-energy initial conditions ($\mathcal{E}<\mathcal{E}_{\rm t}$), showing a strong correlation with the topography. Panel (d): same as (c) for a high-energy initial condition ($\mathcal{E}>\mathcal{E}_t)$, where the correlation with the topography is absent and a large-scale condensate has formed. Visualizations are adapted from \cite{gallet2024two}. \textcolor{black}{Source: Basile Gallet, J. Fluid Mech., 988, A13, 2024; licensed under a Creative Commons Attribution (CC BY) license.}}
    \label{fig:2D_turbulence_topography}
\end{figure}

 In a follow-up study, Gallet\cite{gallet2024two} investigated the case where topography is present only at scales smaller than the size of the domain, unlike Bretherton and Haidvogel or Siegelman and Young, who had assumed topography at all scales. Gallet showed that there is in fact a phase transition at the critical energy $\mathcal{E}=\mathcal{E}_{\rm t}$ between a state without large-scale structure at low $\mathcal{E}$ and an inverse cascade with associated large-scale condensation 
 at high $\mathcal{E}$ in this system (cf. Fig.~\ref{fig:2D_turbulence_topography}), generalizing the selective decay principle to include the large-scale condensate, explaining his findings. Very recent results\cite{zhang2024spectral} on quasi-geostrophic flow over small-scale topography revealed even richer behavior than that uncovered by Gallet, with two threshold values of the topography amplitude and bidirectional energy transfers.

Methods of the type described by Gallet provide a possible avenue for predicting the equilibrated state of forced-dissipative flows based on variational approaches initially designed for conservative systems (selective decay or statistical mechanics). The robustness of such approaches and their predictive skill remain to be further assessed.
\subsection{Instability-driven two-dimensional turbulence}
Sustaining any fluid flow in a stationary state against dissipation requires the injection of energy by a forcing mechanism.  Classical choices include time-independent forcing as in the case of Kolmogorov flow,\cite{arnol1960kolmogorov,meshalkin1961investigation, 
gallet2013two} or a stochastic forcing with a constant energy injection rate.\cite{novikov1965functionals} The latter choice in particular has been widely adopted in numerous studies of forced 2D turbulence.\cite{smith1993bose,boffetta2007energy,laurie2014universal,frishman2018turbulence} Both of these examples, stochastic and time-independent forcing, involve a driving protocol that is imposed independently of the flow state. By contrast, many real flows of interest result from instabilities, for instance of convective, shear or baroclinic types,\cite{chandrasekhar2013hydrodynamic,salmon1980baroclinic,vallis2017atmospheric} which are explicitly system-state dependent. Similarly, models of active fluid flows feature scale-dependent viscosities which can be negative at small scales.\cite{slomka2017geometry,alert2022active}

To investigate the impact of the forcing-flow correlation typical of an instability-type forcing mechanism (but absent for stochastic stirring), following earlier work,\cite{jimenez2007spontaneous} van Kan et al.\cite{van2022spontaneous,van2024vortex} recently studied 2D turbulence driven by a superposition of two forces $\mathbf{f} = \mathbf{f}_\gamma \equiv (1-\gamma)\mathbf{f}_{\rm stoch} +\gamma\mathbf{f}_{\rm inst}$, parameterized by a control parameter $\gamma\in [0,1]$. The first term is a stochastic force injecting a constant power within a fixed (thin, small-scale) band consisting of wave numbers $[k_1,k_2]$, independently of the flow state, while the second term is a linear operator acting on the velocity field which is filtered to retain only the same wave numbers $[k_1,k_2]$ as the stochastic term. To saturate the linear instability resulting from the forcing, a nonlinear damping was included in addition to hyperviscosity.

\begin{figure}
    \centering
    \includegraphics[width=\linewidth]{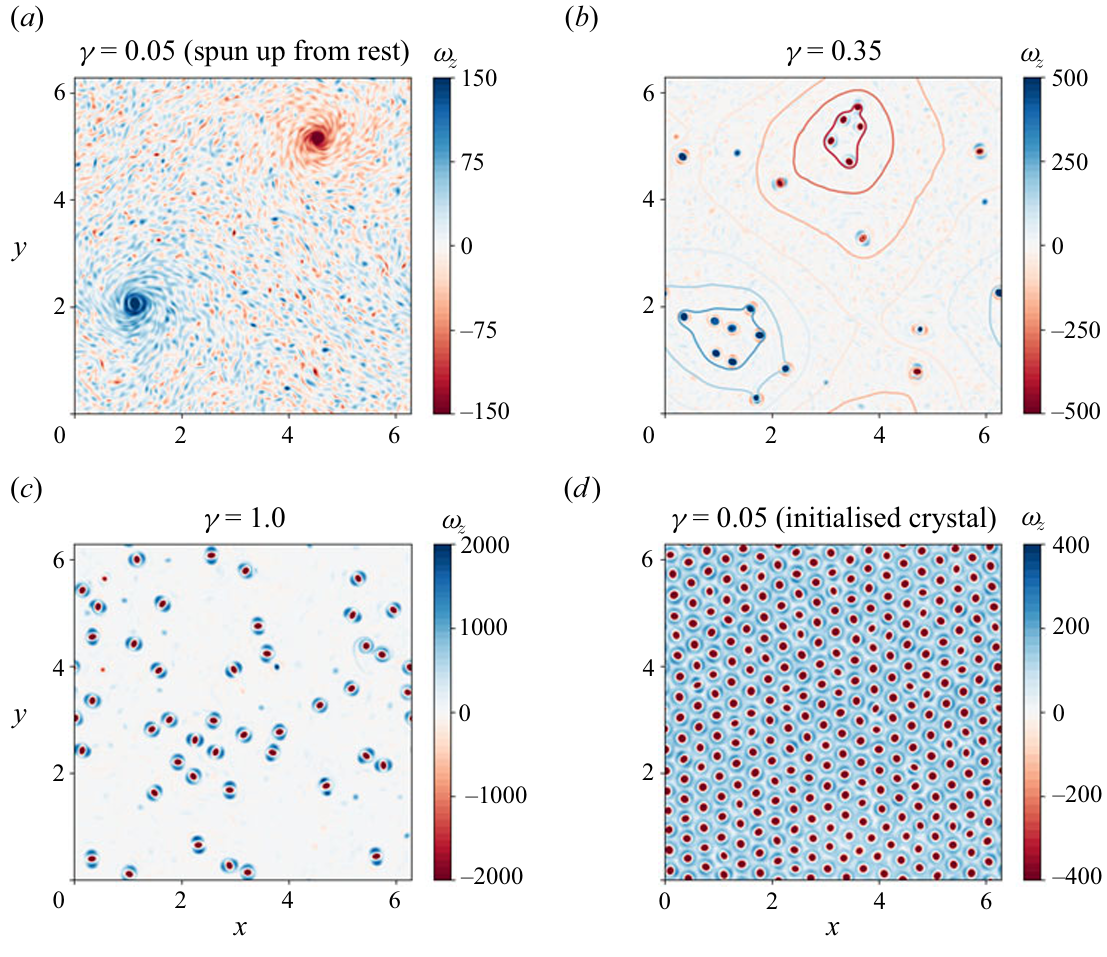}
    \caption{Illustrations (showing $\omega_z$) of different flow states observed in instability-driven 2D turbulence. (a): large-scale condensate, (b) hybrid state with large-scale condensate and embedded shielded vortices of either sign. (c): shielded vortex gas of broken symmetry. (d): vortex crystal. \textcolor{black}{Reproduced with permission from J. Fluid Mech. 380:20210049 (2024). Copyright 2024 Cambridge University Press.}}
\label{fig:inst_driven_turb}
\end{figure}

For small values of $\gamma$, where the random forcing dominates, integrating the equations from rest yields an inverse energy cascade producing a large-scale condensate (Fig.~\ref{fig:inst_driven_turb}(a)). By contrast, as $\gamma$ is increased (stronger instability forcing) the flow morphology changes drastically, and coherent vortices emerge, which are tripolar (consisting of a vorticity extremum in the core and two satellites of the opposite sign at 180$^\circ$ apart). These coherent vortices are also shielded in the sense that the circulation they induce is close to zero outside a finite radius (slightly larger than the forcing scale).  At intermediate values of $\gamma$, shielded vortices of both signs persist, embedded inside the large-scale vortices (Fig.~\ref{fig:inst_driven_turb}(b)). At larger $\gamma$, another phase transition occurs, where the condensate is suppressed and the vortices undergo a spontaneous symmetry breaking, with all vortices of one (arbitrary) sign being annihilated, leading to a polarized vortex gas (Fig.~\ref{fig:inst_driven_turb}(c)). The density of this vortex gas slowly increases in time due to random nucleation, up to a density threshold where an explosive nucleation phase leads to a final state consisting of a high-density gas. This dense vortex gas can in turn be continued to smaller $\gamma$ (weaker instability forcing),  where, at a critical forcing strength $\gamma=\gamma_c$, it forms a hexagonal vortex crystal (Fig.~\ref{fig:inst_driven_turb}(d)), not unlike those found in rotating flows,\cite{boubnov1986experimental,di2020phase} at the Jovian poles,\cite{adriani2018clusters} and in active matter, both in experiments\cite{riedel2005self,xu2024self} and simulations.\cite{james2018turbulence,james2021emergence}. A suitable order parameter for the crystallization transition is the diffusivity of individual shielded vortices, which undergo random Brownian-like motion in the gas phase but are trapped in the crystal phase. The diffusivity shows a power law dependence on $\gamma-\gamma_c$, which has not yet been explained theoretically.

These results highlight the forcing dependence of 2D turbulence, implying non-universality.\cite{linkmann2020non} It is anticipated that the properties of instability-driven turbulence described above can be verified in more complex quasi-2D flows driven by spectrally localized instabilities beyond this simple model. However, many questions remain open, regarding the robustness of the observed phases to changes in control parameters other than $\gamma$, as well as the stability of the shielded vortices in a quasi-2D configuration, and the theoretical description of the different transitions (including crystallization) observed in this forced-dissipative system. Moreover, while many features of inertial instability-driven turbulence at high $\rm Re$ resemble (low-$\rm Re$) active fluid flows, including at the level of the mathematical models (Swift-Hohenberg-Toner-Tu equations\cite{alert2022active}) and the observed flow states, the precise relation between these two physically distinct problems remains to be clarified.

\section{Conclusions\label{sec:conclusions}}
In this mini-review, we have surveyed recent developments in the area of phase transitions in highly anisotropic turbulent flows. We have described how turbulence transitions from 3D to quasi-2D and 2D in different systems. \textcolor{black}{In the presence of large-scale damping, the transition towards the inverse energy cascade is continuous (i.e., second-order) in terms of the order parameter given by the inverse energy flux. The scaling of the inverse energy flux with the physical control parameter near onset of the inverse cascade has been found to be approximately linear in several cases including thin-layer turbulence, but more data is needed and no theory exists to predict the behavior of the system in this parameter regime. When large-scale damping is absent, the transition from small-scale 3D turbulence to the large-scale condensate is discontinuous (i.e., first-order) in terms of the large-scale kinetic energy in the thin-layer and rotating turbulence cases. If the anisotropy is increased sufficiently, e.g., by reducing the layer depth to the Kolmogorov scale, the kinetic energy in 3D modes vanishes with a different power law, whose exponent appears to be larger than one, which also remains incompletely understood. Importantly, nearly all studies summarized above adopted periodic boundary conditions in the horizontal \textit{and} vertical directions. How these transitions are impacted by no-slip boundary conditions, as they are encountered in laboratory experiments, remains yet to be studied. The addition of the competing effects of rotation and stratification enriches the dynamics near the onset of the inverse cascade and more work remains to be done to characterize the large-scale vortices emerging in this case.} We also discussed key examples of multistability and turbulent noise-induced transitions and reviewed recently discovered phase transitions in modified forms of 2D turbulence, \textcolor{black}{where approaches from equilibrium statistical theory have proven successful in some cases, but the limitations of such methods to forced-dissipative turbulence more generally remain incompletely understood.} 

The problems described above are of course highly idealized and pale in comparison to the full complexity of the Earth System. However, they nonetheless pose significant challenges to our understanding and allow us to learn about the fundamental processes of turbulence, \textcolor{black}{including the direction and rate of energy transfers across scales and the associated formation or absence of large-scale flow structures}, which govern not only the transport of heat, momentum, but also pollutants and nutrients on our planet and beyond.  Going forward, the insights gained about geophysical and astrophysical turbulence in the process may help inform parameterizations of turbulence in weather and climate models, which do not typically take inverse energy cascades into account.

Finally, it is interesting to note that the phase transitions discussed here can be viewed as a form of tipping points\cite{lenton2011early} in turbulence, namely, thresholds in parameter space, which, when crossed, induce a rapid, qualitative change in the global system state. \textcolor{black}{As the control parameters (e.g., of the Earth system) changes slowly, the response in the resulting turbulent fluid flows may be sudden.}

\section*{Acknowledgements}
 \textcolor{black}{The author thanks two anonymous referees for their highly constructive reviews which have significantly improved this manuscript.} This work was supported by the National Science Foundation (grants DMS-2009563 and DMS-2308337) and the German Research Foundation (DFG  Projektnummer: 522026592). Part of the writing of this article was done while the author was a Staff Member at the Woods Hole Oceanographic Institution Summer Program in Geophysical Fluid Dynamics summer program 2024, with partial support from NSF (grant DMS-2308338). The author thanks Alexandros Alexakis, Santiago Benavides, Lichuan Xu and Edgar Knobloch for their feedback on earlier versions of this manuscript and Chia-Chin Chung for her support with graphic design. 

\section*{Data Availability Statement }
Data sharing is not applicable to this article as no new data were created or analyzed in this review.

\section*{References}
\hspace{-6.0cm}

\bibliography{aipsamp}
\end{document}